\documentclass[amsfonts,amssymb,twocolumn,showpacs,nofootinbib]{revtex4-1}

\usepackage{amsmath}
\usepackage{amsfonts}
\usepackage{amssymb}
\usepackage{mathrsfs}
\usepackage{amsthm}
\usepackage[colorlinks=true]{hyperref}
\usepackage[all]{hypcap}
\usepackage{marginnote}
\usepackage{graphicx}
\usepackage[usenames]{color}
\usepackage{tikz}
\usepackage{xspace}
\usepackage{verbatim}
\usetikzlibrary{calc}

\usepackage[T1]{fontenc}
\usepackage{microtype}

\newcommand{\nn}{\nonumber}

\newcommand{\cd}{\nabla}
\newcommand{\pd}{\partial}

\newcommand{\txt}[1]{{\textrm{\tiny{#1}}}}
\newcommand{\kin}{\txt{kin}}
\newcommand{\NS}{\txt{NS}}
\newcommand{\BH}{\txt{BH}}
\newcommand{\bin}{\txt{bin}}
\newcommand{\rad}{\txt{rad}}

\newcommand{\WF}{\txt{WF}}
\newcommand{\eff}{\txt{eff}}
\newcommand{\ext}{\txt{ext}}
\newcommand{\INT}{\txt{int}}
\newcommand{\mpl}{m_\txt{pl}}
\newcommand{\calL}{\mathcal{L}}
\newcommand{\calA}{\mathcal{A}}
\newcommand{\plusonetwo}{+\left( 1\leftrightarrow 2 \right)}
\newcommand{\minusonetwo}{-\left( 1\leftrightarrow 2 \right)}
\newcommand{\avg}[1]{\left< #1 \right>}
\newcommand{\lstf}{\langle}
\newcommand{\rstf}{\rangle}
\newcommand{\accR}{\mathscr{R}}
\newcommand{\accS}{\mathscr{S}}
\newcommand{\accW}{\mathscr{W}}

\newcommand{\deform}{\txt{def}}
\newcommand{\ppE}{\txt{ppE}}
\newcommand{\SNR}{\textrm{SNR}}

\newcommand{\lcheck}{\marginnote{$\textcolor{red}{\checkmark}$}}
\renewcommand{\lcheck}{}

\DeclareMathOperator{\sgn}{sgn}

\DeclareSymbolFont{euletters}{U}{eur}{m}{n}
\DeclareMathSymbol{\wp}{\mathord}{euletters}{"7D}

\begin{document}

  \title{Parametrizing and constraining scalar corrections to general relativity}
  \author{Leo C. Stein}
  \thanks{Einstein fellow}
  \email{leostein@astro.cornell.edu}
  \affiliation{Center for Radiophysics and Space Research, Cornell
    University, Ithaca, New York 14853, USA}
  \author{Kent Yagi}
  \email{kyagi@physics.montana.edu}
  \affiliation{Department of Physics, Montana State University,
    Bozeman, Montana 59717, USA}

  \date{\today}

  \begin{abstract}
    We parametrize a large class of corrections to general relativity
    which include a long-ranged gravitational scalar field as
    a dynamical degree of freedom in two ways: parametrizing the
    structure of the correction to the action, and parametrizing the
    scalar hair (multipole structure) that compact objects and
    black holes attain. The presence of this scalar hair violates
    the no-hair theorems present in general relativity,
    which leads to several important effects. The effects we consider
    are (i) the interaction between an isolated body and an external
    scalar field, (ii) the scalar multipole-multipole
    interaction between two bodies in a compact binary, (iii) the
    additional pericenter precession of a binary, (iv) the scalar
    radiation from a binary, and (v) the modification to the
    gravitational wave phase from a binary. We apply this framework to
    example theories including Einstein-dilaton-Gauss-Bonnet gravity
    and dynamical Chern-Simons gravity, and estimate the size of the
    effects.
    Finally, we estimate the bounds that can be placed on parameters of
    the theories from the
    precession of pulsar binaries and from gravitational waves.
  \end{abstract}

 \pacs{04.50.Kd,04.25.Nx}

 \maketitle



\section{Introduction}
\label{sec:intro}

General relativity (GR) is known to be consistent with all experimental and
observational tests to date~\cite{TEGP,lrr-2006-3}. However, all of
these tests are in the weak-field, low-curvature, and low-velocity
regime of the theory. According to the modern paradigm of effective
field theories, we expect GR to require corrections at a
yet-unexplored curvature scale, associated with some new length
scale. In the regime where these corrections are small, they can be
explored by perturbing away from known GR solutions.

One of the most commonly explored extensions of general relativity
is that of a theory with both a metric and a scalar, motivated by
fundamental or effective field theories. In such theories, a
``gravitational'' scalar is (1) long ranged and (2) couples weakly.
Such a theory must reduce to GR in the weak-field, slow-motion limit,
so that the scalar is negligible in the Solar System, and thus evades
weak-field tests. Some examples of such theories are
Brans-Dicke~\cite{TEGP,lrr-2006-3}, Einstein-dilaton-Gauss-Bonnet
(EDGB) gravity~\cite{Moura:2006pz,Pani:2009wy}, and dynamical Chern-Simons
(dCS) gravity~\cite{Jackiw:2003pm,Alexander:2009tp}.
Black holes and neutron stars have been considered in these theories
in several
papers~\cite{Kanti:1995vq,Yunes:2009ch,Yunes:2009gi,2011PhRvD..83j4002Y,Pani:2011gy,Pani:2011xm,Yagi:2012ya,2013PhRvD..87h4058Y}.

A common theme in these theories is that stationary compact objects,
such as neutron stars (NSs) and/or black holes (BHs), act as effective
sources of the long-ranged scalar field, acquiring scalar charge
(called ``hair'' for BHs).
The presence of this scalar charge violates the no-hair theorems of GR
for black holes. For neutron stars the scalar field strength can
depend on the internal structure of the star, which violates the
principle of effacement.\footnote{The effacement principle in GR says that
  at low order, a gravitational body may be described as a point
  particle with the same mass~\cite{1983grr..proc...58D}. In the post-Newtonian (pN) expansion,
  finite-size effects enter at 5pN order~\cite{lrr-2006-4}, while in
  the extreme mass-ratio limit they enter at fourth order in the mass
  ratio~\cite{Galley:2008ih}.} The magnitude of this scalar charge
depends on the size and mass of the body in relation to the new length
scale.

The presence of this scalar field modifies dynamics of a compact
binary system in both the conservative and dissipative
sectors. Consequently, a binary experiences additional pericenter
precession, and the gravitational wave signature is modified. Both of
these effects are observable and therefore can be used to constrain
the coupling strength (which can be quantified through a length $\ell$) of the
scalar interaction in this type of theory.

In this paper we parametrize the types of scalar interactions and the
multipole structure acquired by BHs and NSs. Using this
parametrization we calculate observable effects in a compact binary
system (the additional pericenter precession and modification to
gravitational wave phase). Using these observables and the estimated
scaling laws for scalar multipole moments in this class of theories,
we are able to estimate the bounds which can be placed on the
coupling strength of the nonminimal scalar interaction of these
theories.

We find a specific power-law scaling for the estimated bound on $\ell$
in terms of the binary orbital velocity, with the power determined by
the parametrization of the theory. The pericenter precession bound improves (goes to
shorter lengths) with higher-velocity binaries. Gravitational waves
are estimated to provide an even stronger constraint than pulsar
binaries. Gravitational waves from comparable mass-ratio binaries with
stellar masses are estimated to provide bounds on the order of the
gravitational lengths present in the system: a combination of the
gravitational radii, extents of the bodies, and their separations. For
NS-NS binaries this gives a typical length at the kilometer scale.



The plan for this paper is as follows.
In Sec.~\ref{sec:setup}, we lay out the two parametrizations we use in
this paper.
In Sec.~\ref{sec:multipoles}, we lay out the multipole structure of the
scalar field of an isolated compact object, a compact binary system,
and the scalar radiation field thereof.
In Sec.~\ref{sec:multipole-estimates} we estimate the scalings of
scalar multipole moments of compact objects, which leads to scaling
estimates of scalar multipoles of a compact binary system.
In Sec.~\ref{sec:scalar-interaction} we derive the scalar field-pole
interaction and force on an isolated body, and the scalar pole-pole
interaction and force in a compact binary system.
In Sec.~\ref{sec:pericenter-precession} we compute the additional
pericenter precession in a binary due to the pole-pole interaction.
In Sec.~\ref{sec:rad-reaction} we compute the radiation reaction
due to the flux produced by the binary.
In Sec.~\ref{sec:gw-sig} we compute the modification to the
gravitational wave signature and the parametrized post-Einsteinian
parameters.
In Sec.~\ref{sec:bounds-est} we estimate the bounds which could be
placed from pericenter precession and gravitational wave measurements.
We conclude in Sec.~\ref{sec:conclusion}.

\section{Parametrization of theories}
\label{sec:setup}

Throughout, we will work in units $c=1=\hbar$ where
$[L]=[T]=[M]^{-1}$, and recall that $(8\pi G)^{-1}=\mpl^{2}$
so $[G]=[L]^{2}$.  We take the scalar $\theta$ to have length
dimensions $[\theta]=[L]^{-1}$, so the dimensions of its source are
$[\tau]=[L]^{-3}$. From the structure of the multipolar expansion, an
$s$-pole tensor $\mu^{S}$ has length dimensions $[\mu^{S}]=[L]^{s}$.

We take the full action to be given by the Einstein-Hilbert action for
gravity, a canonical kinetic term for the scalar field, a nonminimal
interaction term between the scalar field and gravity, and a matter
term:
\begin{equation}
\label{eq:action}
S = S_{\txt{EH}} + S_{\kin} + S_{\INT} + S_{\txt{mat}}
\end{equation}
\begin{align}
\label{eq:S-EH}
S_{\txt{EH}} &= \int \frac{1}{2} \mpl^{2} R \sqrt{-g} d^{4}x \\
\label{eq:S-kin}
S_{\kin} &= \int -\frac{1}{2}(\pd_{a}\theta)(\pd^{a}\theta) \sqrt{-g} d^{4}x\\
\label{eq:S-INT}
S_{\INT} &= \int \calL_{\INT}[\theta,g,\epsilon,\cd,R]\sqrt{-g} d^{4}x \,,
\end{align}
where $\cd$ is the Levi-Civita connection of the metric $g$ and
$\epsilon$ its volume form.
A more general action would include a potential for the scalar, but in
order to treat the scalar as gravitational, we take the potential
to be flat so that the scalar is long ranged (including a mass term
would give different phenomenology~\cite{Alsing:2011er,Berti:2012bp}).
We work in the Jordan frame where matter fields couple minimally to the
metric, so the scalar field does not appear in the matter action. This
is appropriate since higher-curvature and higher-derivative actions
may not be conformally transformed to the Einstein
frame~\cite{Bloomfield:2011np}.

We expand the interaction term $\calL_{\INT}$ in powers of
$\theta$. For the purposes of this paper, we are interested in the
linear part: the part with no powers of $\theta$ is not an
interaction, so the linear part leads in the expansion; and it would
require fine-tuning for there to be no linear piece.
In this work we study only the linear-in-$\theta$ interaction.
This may always be written as
\begin{equation}
\calL_{\INT} \sim \theta \ T[g,\epsilon^{0,1},\cd ,R]
\end{equation}
via integration by parts to remove all derivatives from $\theta$. Here
$T$ is a tensor constructed from the metric, zero or one epsilon
tensors,\footnote{\label{fn:epsilon}%
Exactly zero or one epsilon tensors may appear,
  owing to the identity $\epsilon^{i_{1}\cdots
    i_{n}}\epsilon_{j_{1}\cdots j_{n}} = (\sgn{g})n!
  \delta^{[i_{1}}_{j_{1}} \cdots \delta^{i_{n}]}_{j_{n}}$.} the
Levi-Civita connection of the metric and its curvature.
At the linear order, the contribution to the scalar field coming from
different types of operators within $T$ will simply
superimpose. Therefore we will study each type of term separately by
considering only
\begin{equation}
\calL_{\INT} \sim \theta \ T[g,\epsilon^{0,1},\cd^{d} ,R^{r}]
\end{equation}
which is built from $d$ covariant derivatives and $r$ curvature
tensors. By counting
indices, $d$ must be even in four dimensions.

It is clear that we must introduce a new length scale, $\ell$, related
to a cutoff of the effective field theory. This length scale
quantifies the strength of this interaction Lagrangian and the
curvature radius where the interaction becomes important. There are
several ways to parametrize this length (including in terms of a
cutoff---see Appendix~\ref{sec:cutoff}). For future
convenience we choose
\begin{equation}
\label{eq:L-INT}
\calL_{\INT} \sim (\mpl\ell)\, \ell^{\wp} \ \theta \ T[g,\epsilon^{0,1},\cd^{d} ,R^{r}]
\lcheck
\end{equation}
where for dimensional correctness we have
\begin{equation}
\wp = d+2r-3\,.
\lcheck
\end{equation}
Thus the parametrization of the scalar
interaction is given by the integers $(|\epsilon|,d,r)$ (where
$|\epsilon|=0,1$ counts the appearance of $\epsilon$ tensors).

This interaction term is responsible for sourcing the scalar field, in
its equation of motion
\begin{equation}
\label{eq:EOM-theta}
\square\theta = -4\pi \tau
\end{equation}
where clearly $\tau \sim (\mpl\ell) \ell^{\wp} \
\lcheck
T[g,\epsilon^{0,1},\cd^{d},R^{r}]$. This equation must be solved in
the curved strong-field region. However in the far
field, through asymptotic matching, the source term can be replaced
with an effective source term (for example,
see~\cite{2012PhRvD..85f4022Y,2013PhRvD..87h4058Y} and for a more
general discussion see~\cite{Gralla:2013fr}). The far field solution
will be dominated by some lowest nonvanishing scalar multipole moment, which
has a corresponding effective source term which we discuss in
Sec.~\ref{sec:multipoles}.

The nonvanishing scalar moment of lowest multipole order dominates much of the
phenomenology associated with the scalar field. The scalar structure
must be determined on a per-theory basis through asymptotic matching
to a genuine strong-field solution. For the purposes of this paper,
we simply parametrize it by some integers. We will take a NS
to have scalar multipole tensors $\mu_{\NS}^{S}$ with $S$ a multi-index of
valence $s=|S|$, i.e.~$S=k_{1}\cdots k_{s}$. The first few of these
may identically vanish; the lowest nonvanishing one is numbered with
$s=\ell_{\NS}$. Similarly, a black hole has scalar multipole tensors $\mu_{\BH}^{H}$,
and the lowest nonvanishing tensor is numbered with $h=\ell_{\BH}$.

The scalar multipole moments of a compact binary system,
$\mu_{\bin}^{B}$, are determined directly from the moments of the
constituent bodies $\mu_{A}^{I}$ (with $A=1,2$) and their orbits (see
Sec.~\ref{sec:multipoles/binary-radiation}).  In a compact binary
(consisting of BH-BH, BH-NS or NS-NS) the lowest nonvanishing scalar
moment of the binary is simply the lower of the two constituents,
$\ell_{\bin}=\min(\ell_1,\ell_2)$.

The motion of the binary will lead to scalar radiation, now with
radiative multipoles which are determined directly from (time
derivatives of) the binary moments $\mu_{\bin}^{W}$ (see
Sec.~\ref{sec:multipoles/binary-radiation}).  However, the lowest
nonvanishing binary source multipole might not be responsible for the
dominant radiation, for example if there is a conservation law that
protects certain multipoles. Therefore we also parametrize the
\emph{dominant} radiation multipole as $\ell_{\rad}\ge\ell_{\bin}$. In
fact the dominant radiative multipole can differ for BH-BH, BH-NS, and
NS-NS binaries. We will suggest in
Sec.~\ref{sec:multipoles/binary-radiation} that the dominant radiative
multipole is $\ell_{\rad}= 1+\ell_{\bin}=
1+\min(\ell_{1},\ell_{2})$. The parametrization of the multipole
structure of the theory is then given by the integers
$(\ell_{\NS},\ell_{\BH},\ell_{\rad})$ with
$\ell_{\rad}^\txt{HH},\ell_{\rad}^\txt{HS},\ell_{\rad}^\txt{SS}$ for
respectively BH-BH, BH-NS, and NS-NS binaries.

\begin{table}[tb]
\begin{ruledtabular}
\begin{tabular}{rccccccccc}
Theory&$|\epsilon|$&d&r&$\wp$\footnotemark[1]&$\ell_\BH$&$\ell_\NS$&$\ell_{\rad}^\txt{HH}$&$\ell_{\rad}^\txt{HS}$&$\ell_{\rad}^\txt{SS}$\\
\hline
 ``Scalar-tensor'' &0& 0& 1 & -1 & $\cdots$\footnotemark[2] & 0 &$\cdots$\footnotemark[2]&1&1\\
 EDGB &0& 0 & 2 & 1 & 0 & 2\footnotemark[3]&1&1&3\footnotemark[3]\\
 dCS &1& 0& 2 & 1 & 1 & 1 &2&2&2
\end{tabular}
\footnotetext[1]{Not independent, $\wp=2r+d-3$.}
\footnotetext[2]{Black holes have no hair in classical scalar-tensor theories.}
\footnotetext[3]{This is expected but has not yet been calculated.}
\end{ruledtabular}
\caption{
\label{table:params}
Parameters of three example theories with a long-ranged,
weakly-coupling gravitational scalar. The interaction Lagrangians
are seen in Eq.~\eqref{eq:ex-interactions}. The parameters of the
Lagrangian are $(|\epsilon|,d,r)$, and the multipole parameters are
$(\ell_{\BH},\ell_{\NS},\ell_{\rad})$. In this Table we have
$\ell_{\rad}=1+\min(\ell_{1},\ell_{2})$, as suggested in
Sec.~\ref{sec:multipoles/binary-radiation}.}
\end{table}

The full parameters of a theory, then, are given by
$(|\epsilon|,d,r,\ell_{\BH},\ell_{\NS},\ell_{\rad})$.  We give these
parameters for a sampling of theories in Table~\ref{table:params}. The
theories which we highlight are classical ``scalar-tensor'' theories,
dynamical Chern-Simons (dCS), and Einstein-dilaton-Gauss-Bonnet (EDGB). The
leading interaction terms for those theories are
\begin{subequations}
\label{eq:ex-interactions}
\begin{align}
\lcheck
\calL_{\INT}^{\txt{S-T}} &\sim \mpl \ \theta\  R \\
\lcheck
\calL_{\INT}^{\txt{dCS}} &\sim \mpl \ell^{2}\ \theta\ {}^{*}\!RR \\
\lcheck
\calL_{\INT}^{\txt{EDGB}} &\sim \mpl \ell^{2}\ \theta\ {}^{*}\!R{}^{*}\!R
\end{align}
\end{subequations}
where the (left)-dual Riemann tensor is
\begin{equation}
{}^{*}\!R^{abcd} = \frac{1}{2}\epsilon^{abef}R_{ef}{}^{cd}
\end{equation}
(and similarly for the Weyl tensor $C$), the Pontryagin density is
\begin{equation}
{}^{*}\!RR = {}^{*}\!R^{abcd}R_{abcd} = {}^{*}\!C^{abcd}C_{abcd}
\end{equation}
and the four-dimensional Euler (or Gauss-Bonnet) density is
\begin{equation}
{}^{*}\!R{}^{*}\!R = {}^{*}\!R^{abcd}{}^{*}\!R_{abcd}\,.
\end{equation}
Each of these three theories has $\ell_{\rad}=1+\min(\ell_{1},\ell_{2})$.

\section{Multipole moments}
\label{sec:multipoles}

We devote this section to the formalism of the latter half of the
parametrization, that of the scalar multipole structure of isolated
bodies, compact object binaries, and scalar radiation fields. The
formalism uses the machinery of symmetric trace-free tensors;
for references, see
e.g.~\cite{Thorne:1980jl,Blanchet:1986vd,Will:1996jr}.

\subsection{Isolated stationary compact object}
\label{sec:multipoles/NS-BH}
Exterior to an isolated stationary compact object, the scalar field will be
dominated by some lowest nonvanishing multipole moment giving rise to
a solution of the form
\begin{equation}
\label{eq:theta-star-WF}
\theta_{*}^{\WF} = \mu_{*}^{S}\pd_{S} \frac{1}{r_{*}}
\end{equation}
where a subscript or superscript asterisk refers to a property of a star
(or BH), here the scalar field of a body $\theta_{*}$, multipole
moment of a body $\mu_{*}$, and field point distance to a body $r_{*}$.
This solution is to be found from a
numerical or analytic solution in the strong field and extracting the
slowest-decaying behavior in the weak-field (WF). We would like to
approximate this with an
effective point-particle source on flat space-time,\footnote{%
We ignore the background cosmology in which the system may be
embedded, which is considered in
e.g.~\cite{Horbatsch:2011ye,Berti:2013gfa}.
This would be an infrared correction, and should be suppressed by the
ratio of the system's length scales to the cosmological length scale.
} i.e.~to perform an
asymptotic matching. On flat space-time, for the scalar equation of
motion given in
Eq.~\eqref{eq:EOM-theta}, the solution would be found from the Green's
function for $\square$, given by
\begin{equation}
\label{eq:box-greens-function}
\theta(t,\mathbf{x}) = \int_{\mathcal{N}}
\frac{\tau_{\eff}(t-|\mathbf{x}-\mathbf{x}'|, \mathbf{x'})}
{|\mathbf{x}-\mathbf{x}'|} d^{3}x'
\end{equation}
where $\mathcal{N}$ is the intersection of the past light cone of
$(t,\mathbf{x})$ and the source region (see~\cite{Will:1996jr} for
details).

The effective point-particle source which reproduces
Eq.~\eqref{eq:theta-star-WF} is given simply by
\begin{equation}
\label{eq:tau-eff}
\tau_{\eff} = (-)^{s}\mu_{*}^{S} \pd_{S} \delta^{(3)}(\mathbf{x}-\mathbf{x}_{*})\,.
\end{equation}
This can be verified directly by inserting Eq.~\eqref{eq:tau-eff} into
Eq.~\eqref{eq:box-greens-function} and integrating by parts $s$ times.

\subsection{Binary and radiation multipoles}
\label{sec:multipoles/binary-radiation}

The far-zone solution for $\theta$ is given by an expansion of
Eq.~\eqref{eq:box-greens-function} for large $r$, which is given
by~\cite{Will:1996jr}
\begin{align}
\label{eq:theta-FZ-NZ-int}
\theta = \sum_{q=0}^{\infty} \frac{(-)^{q}}{q!}
\left(
\frac{1}{r}\mu^{Q}_{\bin}
\right)_{,Q}
\intertext{where}
\label{eq:multipole-source-integral}
\mu^Q_{\bin}(u) = \int_{\mathcal{M}} \tau_{\eff}(u,\mathbf{x}) x^{Q} d^{3}x
\end{align}
are the source multipoles, $u=t-r$ the retarded time,
$\mathcal{M}$ the constant-$t$ hypersurface intersecting the world
tube, and where $x^{Q}=x^{k_{1}}\cdots x^{k_{q}}$. Here $\tau_{\eff} =
\tau_{\eff,1}+\tau_{\eff,2}$ is the superposition of the effective
source terms for bodies $1$ and $2$ on a Keplerian orbit. Thus
multipole moments of the binary are determined from the multipole
moments of the constituent bodies, $\mu_{A}^{S}$ with $A=1,2$,
at positions $\mathbf{x}_{A}$. Superposing the two effective source terms
[Eq.~\eqref{eq:tau-eff}] and evaluating the source multipole integral
[Eq.~\eqref{eq:multipole-source-integral}] gives
\begin{align}
\mu_{\bin}^{Q} &= \mu_{1}^{S}
\int \delta^{(3)}(\mathbf{x}-\mathbf{x}_{1}) x^{Q}{}_{,S} d^{3}x
\plusonetwo \\
\label{eq:mu-bin-from-BHs}
\mu_{\bin}^{Q}&=
\begin{cases}
\frac{q!}{s!}\mu_{1}^{(k_{1}\cdots k_{s}}x_{1}^{k_{s+1}\cdots
 k_{q})} \plusonetwo & q\ge s \\
0 & \text{otherwise}
\end{cases}
\end{align}
where ${}\plusonetwo$ means to add the same expression with labels 1
and 2 exchanged. This has been written as if $\ell_{1}=\ell_{2}$,
which is valid for BH-BH or NS-NS binaries, but the extension to
$\ell_{1}\ne\ell_{2}$ should be clear.
Evaluating these moments requires the identity
\begin{equation}
x^{Q}{}_{,S} =
\begin{cases}
\frac{q!}{s!} \delta_{S}{}^{(k_{1}\cdots k_{s}} x^{k_{s+1}\cdots k_{q})} &
q\ge s \\
0 & \text{otherwise}
\end{cases}
\end{equation}
where $\delta_{A}{}^{B}=\delta_{a_{1}}{}^{b_{1}}\cdots
\delta_{a_{q}}{}^{b_{q}}$ with $|A|=|B|=q$.
The salient feature here is that the $q^{\mathrm{th}}$ binary source
moment contains $\max(0,q-\ell_{A})$ powers of $x_{A}^{k}$.

In the far zone, the scalar field solution is radiative. The
derivatives in Eq.~\eqref{eq:theta-FZ-NZ-int} act on both $1/r$ and each
$\mu_{\bin}^{Q}(u)$ which depends on retarded time. The solution which
dominates has all derivatives acting on $\mu_{\bin}^{Q}(u)$, since any
derivatives which act on $1/r$ introduce additional powers of
$1/r$. Further, when a spatial derivative acts on a quantity which
depends only on retarded time, we have
\begin{equation}
\label{eq:retarded-time-deriv}
\frac{\pd}{\pd x^{i}}F(u) = -n^{i} \frac{\pd}{\pd t}F(u)
\end{equation}
with $n^{a}$ the unit normal direction vector from the origin to some
field point. Thus, the dominant term is
\begin{equation}
\theta_{\rad} = \sum_{q=\ell_{\bin}}^{\infty} \frac{1}{r} \frac{n^{Q}}{q!}
\ {}^{(q)}\!\mu_{\bin}^{Q}(u)
\end{equation}
where $n^{Q} = n^{k_{1}}\cdots n^{k_{q}}$ and where ${}^{(q)}f =
(\pd/\pd t)^{q} f$.  However, there may be a conservation law or a
suppression for $q=\ell_{\bin}$. Therefore to be slightly more
general, we let the radiation have a lowest nonvanishing moment
$\ell_{\rad}$, so the dominant term is
\begin{equation}
\label{eq:theta-rad}
\theta_{\rad}=\frac{1}{r} \frac{n^{W}}{w!}
\ {}^{(w)}\!\mu_{\bin}^{W}(u)
\end{equation}
where $w=|W|=\ell_{\rad}$.

In particular, focus on the lowest nonvanishing source moment
$\mu_{\bin}^{B}$ with $|B|=\ell_{\bin}=\min(\ell_{1},\ell_{2})$. In
this case, the binary source multipole tensor contains zero powers of
the positions of the bodies; it is simply $\mu_{\bin}^{B}=
\mu_{1}^{B}+\mu_{2}^{B}$ which contains no powers of $x_{A}^{i}$.
Clearly, time derivatives of this moment are
${}^{(n)}\mu_{\bin}^{B}= {}^{(n)}\mu_{1}^{B}
+{}^{(n)}\mu_{2}^{B}$. These time derivatives depend on changes to the
internal structure of the bodies, or at best, if a moment depends on
the spin of a body, on the precession of that spin. Both of the
associated time scales are long compared to the orbital time. In
contrast, consider the next highest moment, $1+\ell_{\bin}$.
In the case of $\ell_{1}=\ell_{2}$, we have for the $1+\ell_{1}$
moment
\begin{equation}
\label{eq:mu-bin-w-equal}
\mu_{\bin}^{aS} = w x_{1}^{(a}\mu_{1}^{S)} \plusonetwo
\end{equation}
with $|S|=\ell_{1}$, whereas for, say, $\ell_1<\ell_{2}$ we have
simply
\begin{equation}
\label{eq:mu-bin-w-unequal}
\mu_{\bin}^{aS} = w x_{1}^{(a}\mu_{1}^{S)}\,.
\end{equation}
This already
contains one power of position vectors, and so it will vary on the
orbital time scale, rather than the precession or radiation-reaction
time scale. For this reason, we suggest that
$\ell_{\rad}=1+\ell_{\bin}$ will be the dominant radiative scalar
moment in most cases.

In this case, as we can see from Eq.~\eqref{eq:theta-rad}, we need to
compute $1+\ell_{1}$ time derivatives of $\mu_{\bin}^{aS}$ from
Eq.~\eqref{eq:mu-bin-w-equal} or Eq.~\eqref{eq:mu-bin-w-unequal} if
respectively $\ell_{1}=\ell_{2}$ or $\ell_{1}<\ell_{2}$.
Let us rewrite this on a Kepler orbit using $\mathbf{x}_{1} = (m_{2}/m) \mathbf{x}_{12},
\ \mathbf{x}_{2}=-(m_{1}/m) \mathbf{x}_{12}$ where $m_{A}$ is the mass
of particle $A$, $m=m_{1}+m_{2}$ is the total mass, and
$\mathbf{x}_{12}=\mathbf{x}_1-\mathbf{x}_{2}$ is the directed relative
separation vector. Then we have
\begin{equation}
\mu_{\bin}^{aS} = w x_{12}^{(a}
\left[
\frac{m_{2}}{m}\mu_{1}^{S)} -
\frac{m_{1}}{m}\mu_{2}^{S)}
\right]\,.
\end{equation}
For future brevity we now define
\begin{equation}
\label{eq:mu-red-def}
\mu_{\txt{red}}^{S} \equiv \left[
\frac{m_{2}}{m}\mu_{1}^{S} -
\frac{m_{1}}{m}\mu_{2}^{S}
\right]
\end{equation}
which is akin to a ``reduced'' moment.

Now consider some number of time derivatives of this tensor. Any time
derivatives acting on $\mu_{A}^{S}$ are, by assumption, suppressed by
the ratio of the orbital time to the precession time. Therefore we can
consider all of the time derivatives acting on $x_{12}^{a}$, which is
simply
\begin{equation}
\frac{\pd}{\pd t} x_{12}^{a} = v_{12}^{a}
\end{equation}
the directed relative velocity vector, $\mathbf{v}_{12} =
\mathbf{v}_{1}-\mathbf{v}_{2}$.
This simplifies for a circular orbit, where
\begin{equation}
\left(
\frac{\pd}{\pd t}
\right)^{2}x_{12}^{a} = -\omega^{2}x_{12}^{a}
= - \frac{Gm}{r_{12}^{3}}x_{12}^{a}
= - \frac{1}{(Gm)^{2}}x_{12}^{a} v^{6}
\end{equation}
where $r_{12}=|\mathbf{x}_{12}|$, $\omega$ is the orbital
angular frequency, and using the leading Kepler relation
($v^{2}=Gm/r_{12}$), where now we
write simply $v$ for $|\mathbf{v}_{12}|$.  Then in the circular,
adiabatic limit, where the time derivative of $r_{12}$ is negligible,
we have
\begin{subequations}
\label{eq:derivs-of-x}
\begin{align}
  \left(
    \frac{\pd}{\pd t}
  \right)^{2j} x_{12}^{a} &=
  \frac{(-)^{j}}{(Gm)^{2j}}x_{12}^{a} v^{6j}\\
  \left(
    \frac{\pd}{\pd t}
  \right)^{2j+1} x_{12}^{a} &=
  \frac{(-)^{j}}{(Gm) ^{2j}}v_{12}^{a} v^{6j}
\end{align}
\end{subequations}
with $j$ a non-negative integer. This gives
\begin{subequations}
\label{eq:derivs-of-mu-bin}
\begin{align}
\label{eq:derivs-of-mu-bin-even}
{}^{(2j)}\mu_{\bin}^{aS} &=
  \frac{(-)^{j}w}{(Gm)^{2j-1}}
\mu_{\txt{red}}^{(S} n_{12}^{a)} v^{6j-2}\\
\label{eq:derivs-of-mu-bin-odd}
{}^{(2j+1)}\mu_{\bin}^{aS} &=
 \frac{(-)^{j}w}{(Gm)^{2j}}
\mu_{\txt{red}}^{(S} v_{12}^{a)} v^{6j}
\end{align}
\end{subequations}
where we have rewritten
$\mathbf{x}_{12}=r_{12}\mathbf{n}_{12}=Gmv^{-2}\mathbf{n}_{12}$ in
order to make all of the velocity dependence explicit.
Each additional time derivative increases the pN order by 1.5 (i.e.~it
introduces three powers of $v$). Naturally here we are interested in
either $2j$ or $2j+1$ being equal to $w=\ell_{\rad}=1+\ell_{1}$,
dependent on whether $\ell_{1}$ is even or odd, which we expect to be
the same parity as $|\epsilon|$ (see
Sec.~\ref{sec:multipole-estimates}). Thus for $|\epsilon|=0$ we expect
to
use Eq.~\eqref{eq:derivs-of-mu-bin-odd} with $2j=\ell_{1}=s$, whereas
for $|\epsilon|=1$ we expect to use
Eq.~\eqref{eq:derivs-of-mu-bin-even} with $2j=1+\ell_{1}=1+s$. This
gives
\begin{equation}
{}^{(w)}\mu_{\bin}^{aS} =
\frac{w}{(Gm)^{s}}
\begin{cases}
(-)^{(1+s)/2} \mu_{\txt{red}}^{(S} n_{12}^{a)} v^{3s+1} \,, & |\epsilon|=1 \\
(-)^{s/2} \ \ \quad \mu_{\txt{red}}^{(S} v_{12}^{a)} v^{3s}  \,, & |\epsilon|=0
\end{cases}
\end{equation}
where we are still specializing to the case of
$\ell_{\rad}=w=1+s=1+\ell_{1}$.

\section{Estimates of multipole moments}
\label{sec:multipole-estimates}

For this parametrization of theories, we can make scaling estimates of
the multipole moments of bodies. This is straightforward for weakly
gravitating bodies where the post-Newtonian approximation holds even
in the interior of the body. It is not strictly true for strongly
gravitating bodies, i.e.~NSs or BHs, but we will boldly extrapolate on
the principle of continuity.
This extrapolation has empirical support with evidence coming from
existing BH~\cite{2011PhRvD..83j4002Y,Yunes:2009gi} and
NS~\cite{2013PhRvD..87h4058Y,Pani:2011xm} calculations.

\subsection{Estimates: Isolated stationary compact objects}
\label{sec:multipole-estimates-scos}

The interaction Lagrangian $\calL_{\INT}$ in Eq.~\eqref{eq:L-INT}
gives rise to a source
term written schematically as
\begin{equation}
\lcheck
\tau \sim (\mpl\ell)\ell^{\wp} \  T[g,\epsilon^{0,1},\cd^{d},R^{r}]\,.
\end{equation}
This will enter into the source moments
\begin{equation}
\mu^{Q} = \int \tau x^{Q} d^{3}x = (\mpl\ell) \ell^{\wp} \int
\lcheck
T[g,\epsilon^{0,1},\cd^{d},R^{r}] x^{Q} d^{3}x\,.
\end{equation}
Since exactly zero or one epsilon tensors may appear (see
footnote~\ref{fn:epsilon}), we will treat
the even (zero epsilons) and odd (one epsilon) cases separately. As we
shall see, these estimates suggest that even theories give rise just
to even scalar multipole moments, and odd theories give rise just to
odd scalar multipole moments.

\subsubsection[Even sources]{Even sources: Zero epsilon tensors}
\label{sec:even-sources}

We make the following estimates: each curvature tensor will go as the
density $G\rho$, and each derivative will introduce a power of $R_{*}^{-1}$
where $R_{*}$ is the radius of the star, which we further approximate as
approximately spherical. This gives
\begin{equation}
\mu^{Q} \sim (\mpl\ell)\ell^{\wp} \frac{1}{R_{*}^{d}} \int (G\rho)^{r} r^{q} n^{Q}
d\Omega r^{2} dr \,.
\lcheck
\end{equation}
The angular integration can be performed with the
identity~\cite{Thorne:1980jl}
\begin{equation}
\label{eq:int-of-ns}
\int n^{Q} d\Omega =
\begin{cases}
\frac{4\pi}{q+1} \delta^{(Q)} & q \text{ even} \\
0 & q \text{ odd}
\end{cases}
\end{equation}
where $\delta^{(Q)}= \delta^{(k_{1}k_{2}} \cdots
\delta^{k_{q-1}k_{q})}$, an isotropic tensor. Here we see that only
even $q$'s are sourced. Continuing with $q$ even, and dropping the
tensorial structure, we have
\begin{equation}
\mu^{Q} \sim (\mpl\ell) \ell^{\wp} \frac{1}{R_{*}^{d}}
(G\rho_{0})^{r} R_{*}^{q+3}
\int_{0}^{1} f^{r} u^{q+2} du
\lcheck
\end{equation}
where $f=\rho/\rho_{0}$ is the dimensionless density profile with
$\rho_{0}$ the central density, and $u=r/R_{*}$ is the dimensionless
radius throughout the star. The final integral is dimensionless and
though it contains information about how centrally concentrated the
density profile is, we will drop it.
Further, we will approximate $\rho_{0}\approx M_{*}/R_{*}^{3}$ and
take the compactness as $C_{*}=GM_{*}/R_{*}$. This gives
\begin{equation}
\label{eq:even-mu-Q-general}
\mu^{Q} \sim
(\mpl\ell)
\left(\frac{\ell}{R_{*}} \right)^{\wp}
C_{*}^{r} R_{*}^{q}\,.
\lcheck
\end{equation}

For black holes, we will replace $R_{*}$ with $GM_{*}$ and hence take the
compactness $C_{*}$ to be 1. Then we will have
\begin{equation}
\label{eq:even-mu-Q-BH}
\mu^{Q} \sim
(\mpl\ell)
\left(\frac{\ell}{GM_{*}} \right)^{\wp}
(GM_{*})^{q}\,.
\lcheck
\end{equation}
Consider for an example the case of EDGB where $d=0$, $r=2$, and
the lowest nonvanishing BH moment is $\ell_{\BH}=0$, a ``scalar
charge.'' We find
\begin{equation}
\mu_{\txt{EDGB}} \sim (\mpl\ell) \frac{\ell}{GM_{*}}\,,
\lcheck
\end{equation}
in agreement with the scaling found in~\cite{2011PhRvD..83j4002Y} once
we identify $\mpl\ell^{2}\sim\alpha_{3}/\beta$.

\subsubsection[Odd sources]{Odd sources: One epsilon tensor}
\label{sec:odd-sources}

For the odd case, we will have one curvature tensor contribute
dominantly the mass current $G\rho v^{i}$, while the remaining $r-1$
curvature tensors dominantly contribute simply a mass density
$G\rho$. We will assume the star undergoes solid body rotation about
axis $\hat{S}^{i}$, so that the rotational velocity within the star
can be written as
\begin{equation}
v^{i} = {}^{(3)}\epsilon^{i}{}_{jk}\hat{S}^{j}n^{k} u v_{\txt{eq}}
\end{equation}
where again $u=r/R_{*}$ and $v_{\txt{eq}}$ is the rotational
velocity at the surface at the equator. Inserting these approximations
into the integral we have
\begin{align}
\lcheck
\mu^{Q} &\sim (\mpl\ell) \ell^{\wp}\frac{1}{R_{*}^{d}}\int \epsilon G\rho v
(G\rho)^{r-1} x^{Q} d^{3}x \\
\label{eq:odd-mu-Q-general-Shat-before-integrating}
\lcheck
&\sim (\mpl\ell)\ell^{\wp}\frac{1}{R_{*}^{d}} (G\rho_{0})^{r} R_{*}^{q+3}
\int \hat{S}^{a} v_{\txt{eq}} u^{q+3} n^{aQ} d\Omega du \\
\label{eq:odd-mu-Q-general-Shat}
\mu^{Q} &\sim
v_\txt{eq}\hat{S}
(\mpl\ell)
\left(\frac{\ell}{R_{*}}\right)^{\wp} C_{*}^{r} R_{*}^{q}\,.
\lcheck
\end{align}
As mentioned earlier, the integral in
Eq.~\eqref{eq:odd-mu-Q-general-Shat-before-integrating} is only
nonvanishing for $q$ odd. From here forward we drop the tensor
structure.

Often times we may want this in terms of the dimensional spin angular
momentum of the body used in relativity,\footnote{In geometric units,
  where $[M]=[L]$, spin angular momentum
  ($\mathbf{S}_{\txt{geom}}=\int \mathbf{r}\times \mathbf{v} dm$) has
  dimensions of $[L]^{2}$, but in the units in this paper this
  definition would be dimensionless (i.e.~angular momentum in units of
  $\hbar$, not appropriate for astrophysics). To convert to the usual
  relativists' convention, we include a factor of $G$,
  $\mathbf{S}=G\mathbf{S}_{\txt{geom}}$. This gives the scaling above.
} $S^{i}$, with $[S^{i}]=[L]^{2}$ and $S^{i}\sim \hat{S}^{i}
v_{\txt{eq}} C_{*} R_{*}^{2}$. The above is rewritten as
\begin{equation}
\label{eq:odd-mu-Q-general}
\mu^{Q} \sim
S
(\mpl\ell)
\left(\frac{\ell}{R_{*}}\right)^{\wp} C_{*}^{r-1} R_{*}^{q-2}\,.
\lcheck
\end{equation}
Again for black holes we can replace $R_{*}$ with $GM_{*}$ and take the
compactness to be 1, giving
\begin{equation}
\label{eq:odd-mu-Q-BH}
\mu^{Q} \sim
S
(\mpl\ell)
\left(\frac{\ell}{GM_{*}}\right)^{\wp} (GM_{*})^{q-2}\,.
\lcheck
\end{equation}
As an example, consider dynamical Chern-Simons, where we have
$d=0$, $r=2$, and the lowest nonvanishing BH moment is $\ell_{\BH}=1$,
a scalar dipole moment. We have
\begin{equation}
\mu^{i}_{\txt{CS}} \sim (\mpl\ell) \ell \frac{S^{i}}{(GM_{*})^{2}}
\lcheck
\end{equation}
which agrees with the scaling found in~\cite{Yunes:2009gi} once we
identify $\mpl\ell^{2}\sim\alpha/\beta$\lcheck{}.

\subsection{Estimates: Binary multipoles}
\label{sec:multipole-estimates-bin}

We may estimate the scaling of $\mu_{\txt{red}}^{S}$ (and therefore,
for the case of $w=\ell_{\rad}=1+s$, also ${}^{(w)}\mu_{\txt{bin}}^{aS}$) by using
the compact object scaling found above in
Sec.~\ref{sec:multipole-estimates-scos}. For simplicity we will only
examine the $|\epsilon|=0$ case. Inserting
Eq.~\eqref{eq:even-mu-Q-general} into Eq.~\eqref{eq:mu-red-def}, we
have
\begin{align}
\mu_{\txt{red}}^{S} &\sim
\left[
\frac{m_{2}}{m}
(\mpl\ell)
\lcheck
\left(
\frac{\ell}{R_{1}}
\right)^{\wp}
C_{1}^{r}R_{1}^{s}
\minusonetwo
\right] \\
\mu_{\txt{red}}^{S} &\sim
\lcheck
\frac{\mpl\ell\ell^{\wp}}{m}
\left[
m_{2} (Gm_{1})^{s-\wp} C_{1}^{\wp+r-s}
\minusonetwo
\right]\,.
\end{align}
Now we will take $C_{A}\sim\mathcal{O}(1)$. This expression is
controlled by the difference $s-\wp$. Though we do not give a general
expression, we can give the scaling for several small integer values
of $s-\wp$. Specifically, we give the scalings for values of
$s-\wp=-1,0,+1,+2$ as follows:
\begin{subequations}
\label{eq:mu-red-s-p-all}
\begin{align}
\label{eq:mu-red-s-p--1}
\lcheck
\mu_{\txt{red}}^{S} &\sim (\mpl\ell)\ell^{s}\frac{\ell}{G\mu} \frac{\delta m}{m} & (s-\wp=-1) \\
\label{eq:mu-red-s-p-0}
\lcheck
\mu_{\txt{red}}^{S} &\sim (\mpl\ell) \ell^{s} \frac{\delta m}{m} & (s-\wp=0) \\
\label{eq:mu-red-s-p-1}
\lcheck
\mu_{\txt{red}}^{S} &\sim (\mpl\ell)\ell^{s} \frac{G\mu}{\ell} & (s-\wp=+1) \\
\label{eq:mu-red-s-p-2}
\lcheck
\mu_{\txt{red}}^{S} &\sim (\mpl\ell) \ell^{s}\frac{G\mu}{\ell}\, \frac{G\delta m}{\ell}  & (s-\wp=+2)
\end{align}
\end{subequations}
where $\delta m = m_{1}-m_{2}$ and $\mu=m_{1}m_{2}/m$ is the reduced
mass. As examples, dCS has $s-\wp=0$, whereas in EDGB, for BHs we have
$s-\wp=-1$, and for NSs we expect $s-\wp=+1$.

\subsection{Regime of validity}
\label{sec:validity}

Using these estimates of the multipole moments of compact objects, we
can estimate the regime of validity of this present analysis. In order
for our analysis to be valid, the correction due to the interaction
term Eq.~\eqref{eq:L-INT} must be small compared to the
Einstein-Hilbert term Eq.~\eqref{eq:S-EH}. At the same time, if the
correction starts to become large then there may be other higher-order
interactions which should have also been included that would
contribute.

A simple way to estimate the regime of validity of the theory is to
analyze the ratio of the interaction Lagrangian to the
Einstein-Hilbert Lagrangian,
\begin{equation}
  \label{eq:chi-def}
  \chi \equiv \frac{(\mpl\ell)\ell^{\wp}\theta
    T[g,\epsilon^{0,1},\cd^{d},R^{r}]}{ \frac{1}{2}\mpl^{2}R} \,.
\end{equation}
We use the scaling for $\theta$, from
Eq.~\eqref{eq:theta-star-WF}, Eq.~\eqref{eq:even-mu-Q-general}, and
taking $r\sim R_{*}$,
\begin{align}
\theta &\sim (2s-1)!!\frac{|\mu^{S}|}{r^{s+1}} \nn\\
\theta &\sim
(2s-1)!!
(\mpl\ell)
\left(
\frac{\ell}{R_{*}}
\right)^{\wp}
C_{*}^{r} R_{*}^{-1}
\,.
\end{align}
Inserting this and taking $\cd\to R_{*}^{-1}$, $R\to G\rho$ gives
\begin{equation}
  \chi\sim\frac{(2s-1)!!(\mpl\ell)^{2}(\ell/R_{*})^{\wp}\ell^{\wp}C_{*}^{r}R_{*}^{-d-1}(G\rho)^{r}}{\rho}
\end{equation}
and after taking $\rho\sim M_{*}/R_{*}^{3}$ and a bit of
simplification,
\begin{equation}
  \label{eq:chi-simplified}
  \chi\sim(2s-1)!!
  \left(
\frac{\ell}{R_{*}}
  \right)^{2\wp+2}
C_{*}^{2r-1}\,.
\end{equation}
Where this ratio is order unity, $\chi\sim 1$, separates the
regime where corrections are small from the regime where corrections
are order unity or larger.

\begin{figure}[tbp]
  \centering
  \includegraphics[width=\columnwidth]{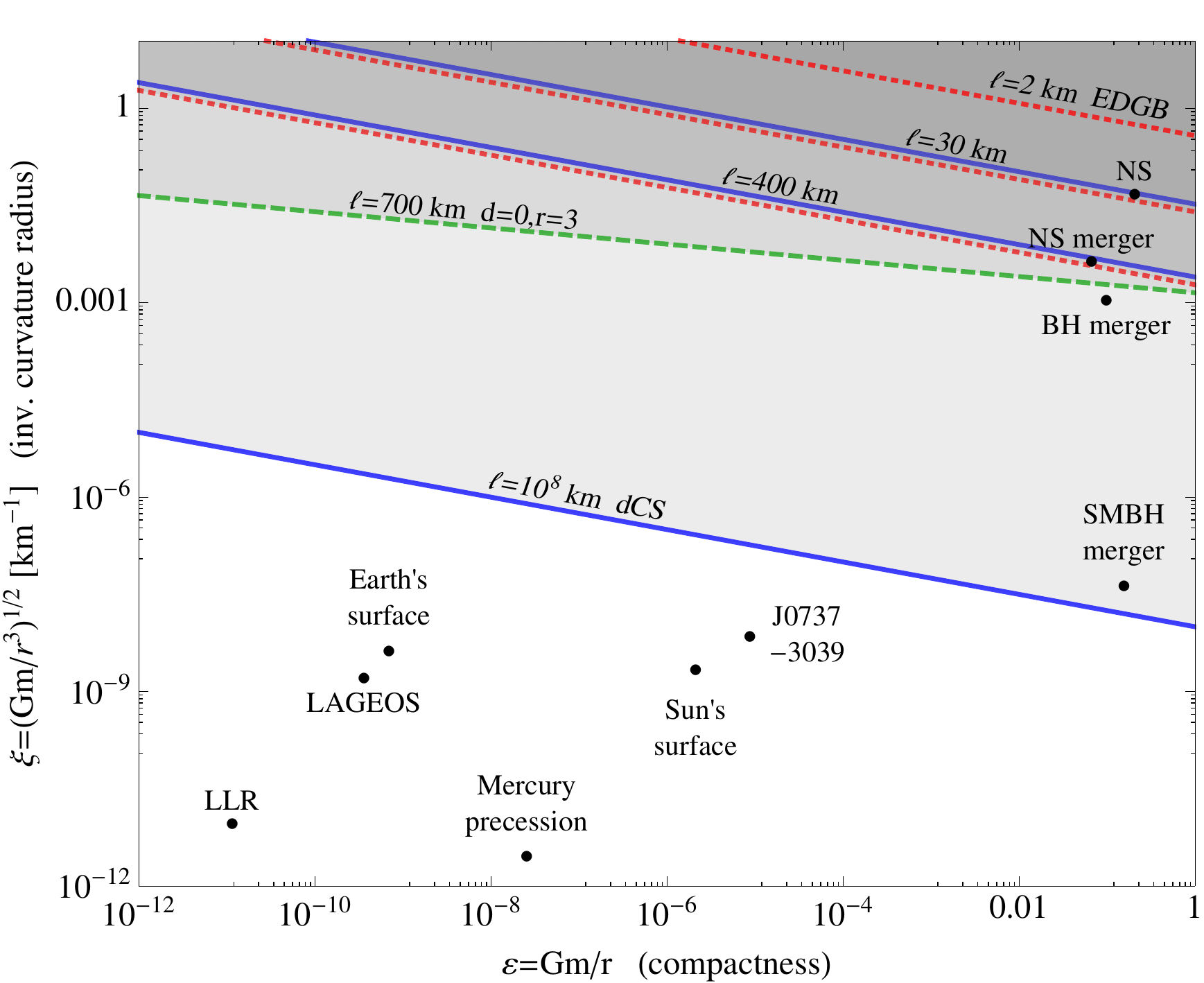}
  \caption{
    Regime of validity (small corrections) and invalidity (large
    corrections) of example theories with different values of $\ell$. The
    shaded region (above the separating line) is the large-correction
    regime. Note it is always the strong-field region (up and to the
    right) which acquires corrections.
    The solid (blue online) curves correspond to dCS, while the dotted
    (red online) curves correspond to EDGB.
    As these two theories have the same $d$ and $r$ parameters, their
    lines have the same slopes [see
    Eq.~\eqref{eq:validity-separatrix}].
    The dashed (green online) curve corresponds to some
    cubic-in-curvature interaction term.
    The vertical axis at left is the inverse curvature radius
    $\xi=(Gm/r^{3})^{1/2}$ in units of km${}^{-1}$. Larger values of
    $\xi$ are stronger fields.
    The horizontal axis gives the dimensionless compactness
    $\varepsilon=(Gm/r)$ of a gravitating system. Larger values of
    $\varepsilon$ are also stronger fields.
    Overplotted are various systems which have been (or will be) used
    as tests of gravity.
    See Appendix~\ref{sec:cutoff} for the relation between $\ell$ and the
    cutoff $\Lambda$.
    \label{fig:validity}}
\end{figure}

In Fig.~\ref{fig:validity} we relate these regimes of validity or
invalidity in a space of compactness vs.~curvature radius,
a choice of parametrization of the space of gravitational phenomena.%
\footnote{This is related to the energy ($E$) vs.~occupation number ($N$)
  characterization more common in EFT. These may be related to more
  geometric quantities and to each other via $|\text{Riem}|^{2}\sim
  \xi^{4}\sim NE^{6}/\mpl^{2}$ and $|\cd\text{Riem}|^{2} \sim
  \xi^{6}/\epsilon \sim NE^{8}/\mpl^{2}$.} On the
horizontal axis we have the dimensionless quantity $\varepsilon\equiv
Gm/r$ where $r$ stands for a typical length scale of some
system, e.g.~orbital radius of a binary or the radius of a neutron
star. Larger $\varepsilon$ is considered deeper into the strong-field
regime. On the vertical axis we have inverse curvature radius
$\xi=(Gm/r^{3})^{1/2}$, with dimensions of km${}^{-1}$. Larger
values of $\xi$ are also considered stronger fields. For reference we
have plotted several gravitational systems which have been used as
tests of gravity in the past or may be used as such in the
future. Some examples of these systems are lunar laser ranging (LLR),
the LAGEOS satellite, the perihelion precession of Mercury, the binary
pulsar system J0737-3039, and the merger of two neutron stars and/or
black holes.

To relate $\chi$ as given in Eq.~\eqref{eq:chi-simplified} to this
plane, we can take $C_{*}\to \varepsilon$, and
$\xi=\sqrt{C_{*}}/R_{*}$ so $R_{*}\to\sqrt{\varepsilon}/\xi$. This
means that the separatrix between small and large corrections is
given on this plane as
\begin{equation}
  \label{eq:validity-separatrix}
  1\sim (2s-1)!!
\varepsilon^{1-d} (\ell\xi)^{2\wp+2}
\end{equation}
for a given $d, r$, and $s$ at a fixed $\ell$.

We plot some examples of separatrices in Fig.~\ref{fig:validity}. For both
EDGB and dCS we have $d=0,r=2$ but we use $s=1$ for dCS and $s=2$ for
EDGB. We have plotted a curve for each theory with both $\ell=30$km
(so the separatrix goes roughly through the NS surface point)
and $\ell=400$km (so the separatrix goes roughly through the NS-NS
merger end point). The shaded region (above each line) is the
large-correction regime, where our analysis is invalid, while the
unshaded region (below each line) is the small-correction regime where
the analysis is valid.
These numbers should be compared to the present bounds. From Solar
System experiments, Ref.~\cite{2011PhRvD..84l4033A} estimated a bound
on $\ell_{\txt{dCS}}\lesssim 10^{8}$km. Meanwhile, from low-mass
x-ray binaries Ref.~\cite{2012PhRvD..86h1504Y} estimated a bound on
$\ell_{\txt{EDGB}}\lesssim 1.9$km.
The lines with the present bounds should be interpreted as follows:
the small-correction regime is at least as large as the unshaded
region shown for each theory, and the strong-correction regime may be
any amount smaller than the shaded region shown.

From Fig.~\ref{fig:validity} it is easy to see why compact binary
systems are a good candidate for testing strong-field corrections to
GR. The top-right corner of this space is generically modified, while
the Solar System and even many binaries are very weakly
modified. NS-NS and stellar mass BH-BH binaries are the dynamical
systems that make it deepest into the strong-field regime (an isolated
NS is also deep in this regime but is not dynamical).

\section{Scalar interaction}
\label{sec:scalar-interaction}

The presence of scalar hair or charge for macroscopic
gravitating bodies leads to a scalar interaction between the field
generated by the body and an external field. This interaction in turn
leads to several effects, among them an additional force on a body, a
change in the binding energy of a compact binary system, and
additional precession of pericenter of a compact binary system. The
simplest approach to computing these effects is to work through an
effective point-particle-with-hair Lagrangian, which is derived by
``integrating out'' the scalar field.
In Sec.~\ref{sec:scalar-general-case} we integrate out the scalar
field for an isolated body in an external scalar field, and find the
scalar force on a body.
In Sec.~\ref{sec:pole-pole} we integrate out the scalar field for a
compact binary system to find the pole-pole interaction, and find the
additional force. This allows for the computation of the additional
pericenter precession (in Sec.~\ref{sec:pericenter-precession}).

\subsection{General case}
\label{sec:scalar-general-case}
Consider an isolated compact object with charge $\mu_{*}^{S}$ giving
rise to $\theta_{*}$ in the weak field, and superpose an external
scalar field $\theta_{\ext}$, $\theta=\theta_{*}+\theta_{\ext}$. The
Lagrangian for the canonical kinetic term, from Eq.~\eqref{eq:S-kin},
becomes
\begin{align}
\calL_{\kin}&= \calL_{\txt{self}} + \calL_{\times} +
\calL_{\ext} \\
\intertext{with}
\calL_{\txt{self}} &=-\frac{1}{2}(\pd_{a}\theta_{*})(\pd^{a}\theta_{*})\\
\calL_{\times} &=-(\pd_{a}\theta_{*})(\pd^{a}\theta_{\ext})\\
\calL_{\ext} &=-\frac{1}{2}(\pd_{a}\theta_{\ext})(\pd^{a}\theta_{\ext})
\end{align}
where $\calL_{\times}$ is the cross term between the field generated
by the body and the external field, and is responsible for the
force.\footnote{The self-term is divergent (it vanishes under
  regularization~\cite{Blanchet:2004bb}), but only the cross term contributes to
  the variation with respect to the location of the body, so the
  self-term may be dropped.} Inserting the isolated WF solution
[Eq.~\eqref{eq:theta-star-WF}] and integrating,
\begin{align}
L_{\times}[\mathbf{x}_{*}]&= \int \calL_{\times} d^{3}x \\
&= -\int
\mu_{*}^{S}\left(\pd_{aS} \frac{1}{r_{*}}\right)
\left(
\pd^{a}\theta_{\ext}
\right) d^{3}x \\
L_{\times}[\mathbf{x}_{*}]&= \int (-)^{s}
\mu_{*}^{S}\left(\pd^{a}\pd_{a} \frac{1}{r_{*}}\right)
\left(
\pd_{S}\theta_{\ext}
\right) d^{3}x\,.
\end{align}
Now with the identity
\begin{equation}
\cd^{2}\frac{1}{r} = -4\pi\delta^{3}(\mathbf{x})
\end{equation}
we easily find
\begin{equation}
L_{\times}[\mathbf{x}_{*}] = (-)^{s+1}4\pi\mu_{*}^{S}
(\pd_{S}\theta_{\ext})[\mathbf{x_{*}}]
\end{equation}
where the derivatives of $\theta_{\ext}$ are evaluated at the location
$\mathbf{x}_{*}$. The effective interaction Lagrangian can be
treated as the negative of an interaction potential. The force on
the body can be found as minus the particle derivative [$\pd_{i}^{(*)}$] of the
potential (i.e.~differentiating with respect to the location of the
particle)~\cite{2012PhRvD..85f4022Y}. We find
\begin{align}
F_{*}^{i}&= \pd^{(*)}_{i} L_{\times}[\mathbf{x}_{*}] \\
F_{*}^{i}&= (-)^{s}4\pi\mu_{*}^{S}
(\pd_{iS}\theta_{\ext})[\mathbf{x_{*}}]
\end{align}
where the extra sign change comes from $\pd_{i}^{(*)} = -\pd_{i}$.

\subsection{Compact binary multipole-multipole interaction}
\label{sec:pole-pole}

We now focus on the scalar interaction in a compact binary system with
particles labeled by $A=1,2$, having lowest nonvanishing scalar
multipole moments $\mu_{1}^{S}$ and $\mu_{2}^{T}$ with
$s=|S|=\ell_{1}$ and $t=|T|=\ell_{2}$ (though the calculation holds
for all moments, not just the lowest ones). Again we take the kinetic
term, now inserting $\theta=\theta_{1}+\theta_{2}$ the superposition
of the two bodies' multipolar fields. As before, we have three terms,
\begin{equation}
\calL_{\kin} = \calL_{\txt{self}-1} + \calL_{\times} + \calL_{\txt{self}-2}
\end{equation}
with
\begin{align}
\calL_{\txt{self}-1} &= -\frac{1}{2}(\pd_a\theta_1)(\pd^a\theta_1) \\
\calL_{\times} &= -(\pd_a\theta_1)(\pd^a\theta_2) \\
\calL_{\txt{self}-2} &= -\frac{1}{2}(\pd_a\theta_2)(\pd^a\theta_2)\,.
\end{align}
As before, only the cross term contributes to the pole-pole
interaction. Inserting the isolated body WF solution
[Eq.~\eqref{eq:theta-star-WF}] into the cross term, we have
\begin{equation}
L_{\times}[\mathbf{x}_{1},\mathbf{x}_{2}] = -\int \mu_{1}^{S}
\left(
\pd_{aS} \frac{1}{r_{1}}
\right)
\mu_{2}^{T}
\left(
\pd^{a}{}_{T} \frac{1}{r_{2}}
\right) d^{3}x\,.
\end{equation}
Here we may integrate by parts to form either expression
\begin{align}
L_{\times}[\mathbf{x}_{1},\mathbf{x}_{2}]&= (-)^{s+1}4\pi\mu_{1}^{S}\mu_{2}^{T}
\left( \pd_{ST}\frac{1}{r_{2}} \right) [\mathbf{x}_{1}] \\
&= (-)^{t+1}4\pi\mu_{1}^{S}\mu_{2}^{T}
\left( \pd_{ST}\frac{1}{r_{1}} \right) [\mathbf{x}_{2}]\,.
\end{align}
This may also be written as
\begin{equation}
\label{eq:L-int-no-derivs}
L_{\times}[\mathbf{x}_{1},\mathbf{x}_{2}] =
(-)^{t+1}4\pi (2s+2t-1)!! \frac{\mu_{1}^{S}\mu_{2}^{T} n_{12}^{\lstf ST\rstf}}{r_{12}^{1+s+t}}\,.
\end{equation}
This may not look symmetric under exchange of particle labels. However
recall that under $1\leftrightarrow 2$, we have $n_{12}\to -n_{12}$,
so the above transforms $(-)^{s+1}n_{12}^{\lstf ST\rstf}\to
(-)^{t+1}n_{12}^{\lstf ST\rstf}$, hence the above expression is indeed
symmetric under exchange of particle labels.

From this interaction Lagrangian we may find the conservative
shift in the binding energy of a binary. As
Eq.~\eqref{eq:L-int-no-derivs} contains no derivatives it is clear that
when performing the Legendre transform to construct the Hamiltonian,
$L_{\times}[\mathbf{x}_{1},\mathbf{x}_{2}]$ will act just as a
potential. Thus, we immediately see
\begin{align}
\label{eq:delta-binding-E}
\delta E_{\txt{bind}} &=-L_{\times}[\mathbf{x}_{1},\mathbf{x}_{2}] \\
\delta E_{\txt{bind}} &=
(-)^{t}4\pi (2s+2t-1)!! \frac{\mu_{1}^{S}\mu_{2}^{T} n_{12}^{\lstf ST\rstf}}{r_{12}^{1+s+t}}\,.
\label{eq:delta-binding-E2}
\end{align}

It is important to note here that when $s=0=t$, for example in the
case of Brans-Dicke theory, this interaction term has the same radial
dependence as the Kepler interaction term,
$L_{\txt{Kep}}=-Gm_{1}m_{2}/r_{12}$. This suggests that the $s=0=t$
case can be cast as a renormalization of Newton's constant. This one
case is qualitatively different from all other possible values of
$s+t$ and must be treated separately.

We may now compute $F_{1}^{i}=\pd^{(1)}_{i}L_{\times}$,
\begin{equation}
\label{eq:pole-pole-force}
F_{1}^{i}=(-)^{t}4\pi (2s+2t+1)!! \frac{\mu_{1}^{S}\mu_{2}^{T}
  n_{12}^{\lstf iST\rstf}}{r_{12}^{2+s+t}} \,.
\end{equation}
The same calculation for $F_{2}^{i}$ shows easily that
$F_{2}^{i}=-F_{1}^{i}$.

\section{Compact~binary pericenter~precession}
\label{sec:pericenter-precession}

We will now compute the primary observable for a compact binary
pulsar under this conservative correction, the precession of the
pericenter of the binary, $\avg{\dot{\omega}}$. This calculation
usually comes with the rate of change of eccentricity $\dot{e}$, rate
of change of semimajor axis $\dot{a}$, rate of change of inclination
$\frac{d}{dt}\iota$, and rate of change of angle of the ascending node
$\dot{\Omega}$. We will only compute $\dot{\omega}$.

We follow Gauss' perturbation
method~\cite{smart,robertson-noonan,TEGP} with the conventions and notation
of~\cite{TEGP}. One must then calculate
the relative perturbing acceleration
$\delta a_{12}^{i}$ (in any convenient coordinate system). This vector is then
decomposed by projecting onto a \emph{time-varying}
orthonormal triad with $e_{1}^{i}=n_{12}^{i}$ and
$e_{2}^{i}=\hat{L}^{i}$ (and $e_{3}=e_{2}\times e_{1}$ so as to complete the
orthonormal triad). In this triad, the components of
$\delta a_{12}^{i}$ are defined as~\cite{TEGP}
\begin{subequations}
\label{eq:accel-decomp}
\begin{align}
\accR &\equiv \delta a_{12}^{i} e_{1,i}\,, \\
\accW &\equiv \delta  a_{12}^{i} e_{2,i}\,, \\
\accS &\equiv \delta a_{12}^{i} e_{3,i}\,,
\end{align}
\end{subequations}
where inner products are taken with a flat Euclidean metric. With
this decomposition, the pericenter of the osculating orbit evolves
secularly as
\begin{align}
\label{eq:dot-omega-inst}
\dot\omega &= - \frac{p \accR}{he}\cos\phi +
\frac{(p+r) \accS }{h e}\sin\phi -\dot\Omega \cos\iota\,, \\
\dot\Omega &= \frac{\accW r}{h}\sin(\omega+\phi) \csc\iota\,.
\end{align}
Here,
\begin{equation}
\label{eq:p-def}
p \equiv a(1-e^2)
\end{equation}
is the semilatus rectum, $r$ and $\phi$ are the quantities related to
the instantaneous orbital elements, given by
\begin{align}
r &\equiv  \frac{p}{1+e \cos\phi}\,, \\
\label{eq:phidot}
r^2 \frac{d \phi}{dt} &\equiv h \equiv \sqrt{G m p}\,,
\end{align}
where $h$ is the orbital angular momentum per unit
mass.  In Eq.~\eqref{eq:dot-omega-inst}, the RHS is to be
orbit averaged as any quantity $Q$,
\begin{equation}
\label{eq:orbit-ave-def}
\avg{Q} = \frac{1}{T}\oint Q dt = \frac{1}{T} \int_0^{2\pi}
\frac{Q(\phi) \mathrm{d}\phi}{\dot{\phi}}\,,
\end{equation}
where $T$ is the background orbital period $T=2\pi a^{3/2}/\sqrt{G m}$,
$\phi$ is the orbital phase, and the Jacobian $\dot\phi$ must of
course be included [from Eq.~\eqref{eq:phidot}].
This procedure is appropriate when the time derivatives of the
osculating elements are much smaller than the orbital timescale and
there are no resonances. Of course, all of the $\phi$ dependence in
$\accR, \accW, \accS$ and $r$ must be included in the orbit averaging.
This procedure gives, for the leading GR pericenter precession,
\begin{equation}
\label{eq:peri-prec-GR}
\avg{\dot{\omega}}_{\txt{GR}} =
\frac{1}{T}\frac{6\pi Gm}{a(1-e^{2})}
= \frac{3 (Gm)^{3/2}}{(1-e^{2})a^{5/2}}\,.
\end{equation}

With the pole-pole force in hand from Eq.~\eqref{eq:pole-pole-force}
we can compute the relative acceleration,
\begin{equation}
a_{12}^{i} = a_{1}^{i}-a_{2}^{i} =
\left(
\frac{1}{m_{1}}+\frac{1}{m_{2}}
\right) F_{1}^{i} = \frac{1}{\mu}F_{1}^{i}
\end{equation}
where again the reduced mass is $\mu=m_{1}m_{2}/m$.
This acceleration is decomposed as
\begin{subequations}
\label{eq:acc-decomp}
\begin{align}
\accR &= \calA \mu_{1}^{S}\mu_{2}^{T}
\times
n_{12}^{\lstf iST\rstf} n_{12}^{i}
(1+e\cos\phi)^{2+s+t}\\
\accS &= \calA \mu_{1}^{S}\mu_{2}^{T}
\hat{L}^{i}
\times
n_{12}^{\lstf iST\rstf}
(1+e\cos\phi)^{2+s+t} \\
\accW &= \calA \mu_{1}^{S}\mu_{2}^{T}
\epsilon_{ijk}\hat{L}^{j}
\times
n_{12}^{\lstf iST\rstf}n_{12}^{k}
(1+e\cos\phi)^{2+s+t}
\end{align}
\end{subequations}
with all of the $\phi$ dependence to the right of $\times$ in each
expression (remember that $n_{12}$ rotates with the orbit), and where
we have defined
\begin{equation}
\calA \equiv \frac{1}{\mu}(-)^{t}4\pi
(2s+2t+1)!! p^{-(2+s+t)}\,.
\end{equation}
These expressions are subject to the identities~\cite{Blanchet:1986vd}
\begin{equation}
n_{i}n_{\lstf a_{1}\ldots a_{l}\rstf} = n_{\lstf ia_{1}\ldots a_{l}\rstf} +
\frac{l}{2l+1} \delta_{i\lstf a_{1}}n_{a_{2}\ldots a_{l}\rstf}
\end{equation}
and by contracting,
\begin{equation}
n_{i}n_{\lstf iL\rstf} = \frac{l+1}{2l+1}n_{\lstf L\rstf}\,.
\end{equation}

Inserting the decomposed acceleration [Eq.~\eqref{eq:acc-decomp}] into
the expression for $\dot\omega$ [Eq.~\eqref{eq:dot-omega-inst}] and
then orbit averaging [Eq.~\eqref{eq:orbit-ave-def}] we arrive at
\begin{multline}
\label{eq:avg-omega-dot}
\avg{\dot{\omega}} =
\frac{1}{T}\frac{p^{2}}{Gm}\calA\mu_{1}^{S}\mu_{2}^{T}
\left[
-\frac{1}{e} \frac{s+t+1}{2s+2t+1} I_{1}^{ST}
+\frac{1}{e} \hat{L}^{i} I_{2}^{iST}\right.\\
\left.
-\cot\iota \frac{s+t+1}{2s+2t+3} \epsilon_{ijk}\hat{L}^{j} I_{3}^{ikST}
\right]
\end{multline}
where we have defined the three tensor-valued integrals
\begin{subequations}
\begin{align}
I_{1}^{ST} &= \int_{0}^{2\pi} n_{12}^{\lstf ST\rstf}
(1+e\cos\phi)^{s+t}
\cos\phi d\phi\\
I_{2}^{iST} &= \int_{0}^{2\pi}n_{12}^{\lstf iST\rstf}
(1+e\cos\phi)^{s+t-1}
(2+e\cos\phi)\sin\phi d\phi\\
I_{3}^{ikST} &= \int_{0}^{2\pi}\delta^{k\lstf i}n_{12}^{ST\rstf}
(1+e\cos\phi)^{s+t-1}
\sin(\omega+\phi) d\phi
\end{align}
\end{subequations}
which are all functions of eccentricity of order unity.

Here we can immediately extract the relative pN order of this
effect. Recalling that $T\propto a^{3/2}$, $p\propto a$, $a\propto
v^{-2}$, and
$\calA\propto p^{-(2+s+t)}$, we have
\begin{align}
\avg{\dot{\omega}} &\propto v^{2(s+t)+3}\\
\frac{\avg{\dot{\omega}}}{\avg{\dot{\omega}}_{\txt{GR}}}
& \propto v^{2(s+t-1)}\,.
\end{align}
We will go into more detail in the next subsection.
We remind the reader here that for the special case of $s=0=t$, the
pole-pole interaction term can be absorbed by a rescaling of Newton's
constant, so it is not actually pre-Newtonian. For other values of
$s,t$, only the sum $s+t$ enters this scaling, and the pericenter
precession is of relative $+(s+t-1)$~pN order. Remember that $s,t$
should have the same parity, so $s+t$ is even, and therefore this
effect is always of odd relative pN order, starting at relative +1~pN.

\subsection{Pericenter precession scaling estimates}
\label{sec:precession-estimate}
In order to estimate the bounds which may be placed on $\ell$ in a
given theory, we must study how the ratio
$\avg{\dot\omega}/\avg{\dot\omega}_{\txt{GR}}$ scales with $\ell$, $\wp$,
the constituent masses, radii, scalar multipole moments, and
orbital velocity.

The dependence on $\ell$ is buried inside the scaling of the multipole
tensors $\mu_{A}^{Q}$ where $A=1,2$ and $|Q|=s,t$ for respectively
bodies 1,2. We here repeat the scaling found in
Sec.~\ref{sec:multipole-estimates-scos} for $|\epsilon|=0$ (for
$|\epsilon|=1$, the quantity $v_{\txt{eq}}$ must also be included). We
found
\begin{equation*}
\mu^{Q} \sim
(\mpl\ell)
\left(\frac{\ell}{R_{*}} \right)^{\wp}
C_{*}^{r} R_{*}^{q}
\lcheck
\end{equation*}
for a body with radius $R_{*}$ and compactness $C_*$. This we insert
into Eq.~\eqref{eq:avg-omega-dot}. At the same time, we also pull the
eccentricity dependence (in $I_{1,2,3}$) into a function $f_{1}(e)$
which is of order unity. This gives
\begin{multline}
\lcheck
\avg{\dot\omega}\sim(-)^{t}(2s+2t+1)!!
\frac{1}{T}\frac{\ell^{2}}{(Gm)(G\mu)}
\left(
\frac{\ell^{2}}{R_{1}R_{2}}
\right)^\wp
\\
\times
(C_{1}C_{2})^{r}
\left(
\frac{R_{1}}{p}
\right)^{s}
\left(
\frac{R_{2}}{p}
\right)^{t}
f_{1}(e)
\end{multline}
where $R_{A}$ and $C_{A}$ with $A=1,2$ are respectively the radius and
compactness of body $A$. We compare $\avg{\dot\omega}$ to the GR
expression in Eq.~\eqref{eq:peri-prec-GR} by taking their ratio. There
is yet more eccentricity dependence in $p=a(1-e^{2})$ [from
Eq.~\eqref{eq:p-def}] which we absorb into a new function $f_{2}(e)$
which is also of order unity. We will also remove all of the
dependence on the semimajor axis $a$ in favor of the orbital velocity
through the Kepler relation $v^{2}=Gm/a$. This gives
\begin{align}
\label{eq:avg-dot-omega-scaling}
\frac{\avg{\dot\omega}}{\avg{\dot\omega}_{\txt{GR}}}
\sim{}&
(-)^{t}(2s+2t+1)!!\frac{\ell^{2}}{(Gm)(G\mu)}
\left(
\frac{\ell^{2}}{R_{1}R_{2}}
\right)^\wp
(C_{1}C_{2})^{r} \nn\\
&{}\times
\left(
\frac{R_{1}}{Gm}
\right)^{s}
\left(
\frac{R_{2}}{Gm}
\right)^{t}
f_{2}(e)
v^{2(s+t-1)}\,.
\lcheck
\end{align}
This result reproduces the dCS pericenter precession as calculated in
Eq.~(131) of~\cite{2013PhRvD..87h4058Y}.
However, we cannot compare to the EDGB result:
Again we have the caveat that for $s=t=0$, the pole-pole interaction
can be absorbed by rescaling $G$,
so the modification is pushed to a higher order.

Finally, let us note an effect which we have not computed here. There
will be metric deformations which also contribute to pericenter
precession with the same dependence on $\ell$ as this scalar
interaction effect. For example, in dCS the correction to the metric
quadrupole-monopole interaction can dominate over the scalar
dipole-dipole interaction, depending on the spins of the two
bodies~\cite{2013PhRvD..87h4058Y}.

\section{Compact~binary radiation~reaction}
\label{sec:rad-reaction}

In the dissipative sector of the dynamics, the binding energy (and
angular momentum) of the binary is carried away by radiation in all
dynamical fields. Both the metric and scalar contribute, as do any
other additional degrees of freedom, but here we only consider the
scalar field.

The energy flux in some field $\varphi$ is quantified in the
stress-energy tensor\footnote{In the case of the metric, the flux is
  quantified via the effective stress-energy tensor of gravitational
  waves~\cite{Stein:2010pn}.}
$T^{(\varphi)}_{ab}$. Specifically, the energy flux is calculated as
the integral of the flux density over a 2-sphere approaching
asymptotic infinity, captured in the component
$T^{(\varphi)}_{ti}n^{i}$ where $n^{i}$ is the outward unit
normal. We have for such a field (see Sec.~VI
of~\cite{2012PhRvD..85f4022Y})
\begin{equation}
\label{eq:Edotdefinition}
\dot{E}^{(\varphi)} = \lim_{r\to\infty} \int_{S^2_r}
\avg{T^{(\varphi)}_{ti} n^i }  r^2 d\Omega\,,
\end{equation}
where again $\avg{}$ is an orbit-averaged quantity.

For the scalar field with canonical kinetic term and flat potential,
we have the stress-energy tensor
\begin{equation}
\label{eq:Tab-theta}
T_{ab}^{(\theta)} = (\pd_{a}\theta)(\pd_{b}\theta)
- \frac{1}{2} g_{ab} (\pd\theta)^{2}\,.
\end{equation}
Here we will insert the far-zone, radiative solution from
Eq.~\eqref{eq:theta-rad}, and use the identity
[Eq.~\eqref{eq:retarded-time-deriv}] $\pd_{i}\theta =
-n^{i}\pd_{t}\theta$ in the far zone. Combining, we have
\begin{equation}
\label{eq:T-theta-t-ni}
T_{ti}^{(\theta)}n^{i} = -(\pd_{t}\theta)^{2} =
-\frac{1}{r^{2}}\frac{1}{(w!)^{2}}
\left[ n_{W} \  {}^{(w+1)}\!\mu_{\bin}^{W} \right]^{2}
\end{equation}
where $w=|W|=\ell_{\rad}$. The quantity $\mu_{\bin}$ has no dependence
on $n^{i}$, being defined in the near zone. Therefore the angular integral
can be performed, with the aid of Eq.~\eqref{eq:int-of-ns}, giving
\begin{equation}
\dot{E}^{(\theta)} = - \frac{1}{(w!)^{2}} \frac{4\pi}{2w+1}
\delta_{(VW)} \avg{ {}^{(w+1)}\!\mu_{\bin}^{V}
  {}^{(w+1)}\!\mu_{\bin}^{W} }
\end{equation}
where $|V|=w=|W|$ with $V=a_{1}\cdots a_{w}$, $W=b_{1}\cdots b_{w}$,
and $\delta_{(VW)}=\delta_{(a_{1}b_{1}}\cdots\delta_{a_{w}b_{w})}$.

Now we use the calculation of an arbitrary number of derivatives of
$\mu_{\bin}^{aS}$, given in Eq.~\eqref{eq:derivs-of-mu-bin}. Notice
that $1+w=2+s$ has the same parity as $s$. For $|\epsilon|=0$, $s$ is
even and we will take an even number of derivatives, using
Eq.~\eqref{eq:derivs-of-mu-bin-even} with $2j=1+w$, whereas for
$|\epsilon|=1$ we will have $s$ odd, take an odd number of
derivatives, and therefore use Eq.~\eqref{eq:derivs-of-mu-bin-odd}
with $2j=w$. This gives
\begin{equation}
{}^{(w+1)}\mu_{\bin}^{aS} =
\frac{w}{(Gm)^{w}}
\begin{cases}
(-)^{(2+s)/2} \mu_{\txt{red}}^{(S} n_{12}^{a)} v^{3s+4} \,, & |\epsilon|=0 \\
(-)^{(1+s)/2} \mu_{\txt{red}}^{(S} v_{12}^{a)} v^{3s+3}  \,, & |\epsilon|=1\,.
\end{cases}
\end{equation}
Combining, we find
\begin{widetext}
\begin{equation}
\label{eq:Edot-theta}
\dot{E}^{(\theta)} = - \frac{1}{(s!)^{2}} \frac{4\pi}{2w+1}
\frac{1}{(Gm)^{2w}}\mu_{\txt{red}}^{S}\mu_{\txt{red}}^{T}
\delta_{(abST)}
\begin{cases}
\avg{ n_{12}^{a} n_{12}^{b} v^{6s+8} }\,, & |\epsilon|=0 \\
\avg{ v_{12}^{a} v_{12}^{b} v^{6s+6} }\,, & |\epsilon|=1
\end{cases}
\end{equation}
\end{widetext}
where $|S|=s=|T|$ and where we have taken $\mu_{\txt{red}}$ to be
constant over the time scale of an orbit.
Again remember that this is for the case of $w=1+s$.

This is to be compared with the gravitational wave luminosity in GR,
which in the circular limit is given by the well-known expression
\begin{equation}
\label{eq:Edot-GW}
\dot{E}^{\txt{GW}} = -\frac{32}{5} G \mu^{2} r_{12}^{4} \omega^{6} = -\frac{32}{5}\frac{1}{G}\frac{\mu^{2}}{m^{2}}v^{10}\,,
\end{equation}
where in the second equality we have used the Kepler relation.
From this we see that the scalar flux $\dot{E}^{(\theta)}$ is of
relative $+(3s-1)$~pN order compared to the gravitational wave flux.

\subsection{Radiation reaction scaling estimates}
\label{sec:rad-reaction-scaling}
By using the scaling estimates given in
Sec.~\ref{sec:multipole-estimates-bin}, we can estimate the scaling of
the extra energy lost from the binary due to scalar radiation. We are
interested in the (small) ratio $\dot{E}^{(\theta)}/\dot{E}^{\txt{GW}}$, for
the total energy loss is $\dot{E} = \dot{E}^{\txt{GW}}
(1+\dot{E}^{(\theta)}/\dot{E}^{\txt{GW}})$. The flux ratio is
proportional to a power of $\ell$ through $\mu_{\txt{red}}$.

For simplicity we will only consider $|\epsilon|=0$. Combining
Eqs.~\eqref{eq:Edot-theta} and \eqref{eq:Edot-GW}, after a small
amount of algebra, we have
\begin{equation}
\frac{\dot{E}^{(\theta)}}{\dot{E}^{\txt{GW}}} \sim \frac{5\pi^{2}}{(2w+1)(s!)^{2}}
\frac{\mpl^{2}}{\mu^{2}}
\frac{|\mu_{\txt{red}}^{S}|^{2}}{(Gm)^{2s}}
v^{6s-2}\,.
\end{equation}
Now we insert, for $\mu_{\txt{red}}^{S}$, the four scalings we found
in Sec.~\ref{sec:multipole-estimates-bin}, given by the four
differences $s-\wp=-1,0,+1,+2$.
We find, after a bit of simplification,
\newcommand{\thisLHS}{\frac{\dot{E}^{(\theta)}}{\dot{E}^{\txt{GW}}}}
\newcommand{\thisfactor}[1]{\frac{ #1 v^{6s-2}}{(2w+1)(s!)^2} \left( \frac{\ell}{Gm} \right)^{2+2\wp}}
\begin{subequations}
\begin{align}
\lcheck
\label{eq:E-dot-theta-ratio-s-p--1}
\thisLHS &\sim \thisfactor{\eta^{-4}}
\frac{\delta m^{2}}{m ^{2}} & (s-\wp=-1) \\
\lcheck
\label{eq:E-dot-theta-ratio-s-p-0}
\thisLHS &\sim \thisfactor{\eta^{-2}}
\frac{\delta m^{2}}{m^{2}} & (s-\wp=0) \\
\lcheck
\label{eq:E-dot-theta-ratio-s-p-1}
\thisLHS &\sim \thisfactor{}
& (s-\wp=+1) \\
\lcheck
\label{eq:E-dot-theta-ratio-s-p-2}
\thisLHS &\sim \thisfactor{}
\frac{\delta m^{2}}{m^{2}} & (s-\wp=+2)
\end{align}
\end{subequations}
where $\eta=m_{1}m_{2}/m^{2}=\mu/m$.
Some of these expressions reproduce results published previously in
the literature, while others are new. For EDGB with two BHs, we have
$s-\wp=-1$, given in Eq.~\eqref{eq:E-dot-theta-ratio-s-p--1}. This
equation reproduces the same scaling with $\eta$,
the ratio $(\delta m/m)$, and the relative
post-Newtonian order as given by~\cite{2012PhRvD..85f4022Y} in their
Eq.~(134). For dCS and either NSs or BHs, we have $s-\wp=0$, given in
Eq.~\eqref{eq:E-dot-theta-ratio-s-p-0}. This reproduces the scaling
with $\eta$ and the post-Newtonian order given
by~\cite{2012PhRvD..85f4022Y} in their Eq.~(139) (though this
comparison is not well justified, since here we have only estimated
the scaling for $|\epsilon|=0$ theories).

\section{Gravitational wave signature}
\label{sec:gw-sig}

In this section, we derive a correction to the gravitational waveform
phase (in the Fourier domain) of a compact binary system. We only show
the scaling estimate and neglect numerical factors. For simplicity, we
restrict our attention to binaries with a circular orbit, for the case
of $s=t$, and with $|\epsilon|=0$. In order to accomplish this goal,
we need to combine three ingredients: corrections to (i) the binding
energy, (ii) the Kepler relation, and (iii) the energy flux. We then proceed
through the stationary phase approximation (see
e.g.~\cite{Tichy:1999pv}).

In Eq.~\eqref{eq:delta-binding-E2}, we derived a correction to the
binding energy due to the pole-pole interaction. In general, there is
also a correction to the binding energy due to the fact that the
spacetime around the compact object is deformed, which we have not
addressed in this paper. Let us parametrize this by
\begin{equation}
\label{E-deform}
\delta E_\txt{bind}^\deform = \frac{C_\deform}{r^{1+n_\deform}}\,,
\end{equation}
where the coefficient $C_\deform$ has units of
$[C_\deform]=[L]^{n_\deform}$.  Note that this deformation is of
$+n_\deform$~pN order relative to GR. For example, $n_\deform = 2$ for both
EDGB~\cite{2011PhRvD..83j4002Y} and
dCS~\cite{2013PhRvD..87h4058Y,Yagi:2012ya,Yagi:2012vf}. By combining
this with the force due to the pole-pole interaction shown in
Eq.~\eqref{eq:pole-pole-force}, we can derive the equation of motion
of a binary as
\begin{equation}
r_{12} \omega^2 \sim \frac{G m}{r_{12}^2}
\left[
1 + \frac{1}{Gm\mu}\frac{|\mu_{1}\mu_{2}|}{r^{2s}}
+ \frac{1}{Gm\mu} \frac{C_{\deform}}{r^{n_{\deform}}}
\right] \,.
\end{equation}
By assuming that the orbital velocity $v \sim (G m \omega)^{1/3}$ is
much less than the speed of light, we can invert the above expression
by expanding in terms of $v$ and obtain the modified Kepler relation
$r_{12}(\omega)$ as
\begin{multline}
\label{eq:mod-Kepler}
 r_{12}(\omega) \sim \frac{G m}{(G m \omega)^{2/3}}
\left[
1 + \frac{1}{Gm\mu}\frac{|\mu_{1}\mu_{2}|}{(Gm)^{2s}}
(Gm\omega)^{4s/3}\right. \\
\left.
+\frac{1}{Gm\mu}
\frac{C_{\deform}}{(Gm)^{n_{\deform}}} (Gm\omega)^{2n_{\deform}/3}
\right] \,.
\end{multline}
By using Eqs.~\eqref{eq:delta-binding-E2}, \eqref{E-deform}
and~\eqref{eq:mod-Kepler}, we obtain the binding energy in terms of $(G
m \omega)$ as
\begin{multline}
\label{eq:mod-energy}
E_{\txt{bind}}  \sim \mu (G m \omega)^{2/3}
\left[
1 +
\frac{1}{Gm\mu}\frac{|\mu_{1}\mu_{2}|}{(Gm)^{2s}}
(Gm\omega)^{4s/3}\right. \\
\left.
+\frac{1}{Gm\mu}
\frac{C_{\deform}}{(Gm)^{n_{\deform}}} (Gm\omega)^{2n_{\deform}/3}
\right] \,.
\end{multline}

Next, we move onto the dissipative correction, namely the one to the
energy flux. In Eq.~\eqref{eq:Edot-theta}, we derived the energy flux
for the scalar radiation. There is also a correction to the energy
flux for the gravitational radiation that we have not addressed in
this paper. This we parametrize as
\begin{equation}
\dot{E}^{(h)} = C_h \left( \frac{G m}{r_{12}} \right)^{5+n_h}\,,
\end{equation}
where $C_h$ has dimensions of $[C_h] = [L]^{-2}$. Notice that
$\dot{E}^{(h)}$ gives an $+n_h$~pN correction relative to GR. For
example, we have $n_h=0$ for EDGB and $n_h=2$ for
dCS~\cite{2012PhRvD..85f4022Y}.  There are two sources for this
correction. The first is the appearance of $T^{(\theta)}_{ab}$
[Eq.~\eqref{eq:Tab-theta}] on the RHS of the Einstein equations, and
the second is the modification to the LHS of the Einstein equations
(due to $\delta\calL_{\INT}/\delta g$, e.g.~the $C$-tensor in
dCS~\cite{Alexander:2009tp}). Though it is possible to robustly
estimate the scaling due to the stress-energy tensor of the scalar
field, naive scaling estimates for $\delta\calL_\INT/\delta g$ may
fail (as in the case with topological invariants, i.e.~dCS and
EDGB). Without knowing this scaling, we cannot even know which of the
two effects dominate. Therefore we only leave this correction in
terms of the parameters $n_{h}$ and $C_{h}$.

Combining with Eqs.~\eqref{eq:Edot-theta} and~\eqref{eq:mod-Kepler},
we find
\begin{align}
\label{eq:mod-Edot}
\dot{E}  \sim \frac{\eta^2}{G} (G m \omega)^{10/3}
&\left[
1 +
\frac{1}{Gm\mu}\frac{|\mu_{1}\mu_{2}|}{(Gm)^{2s}}
(Gm\omega)^{4s/3}\right. \nn\\
&+\frac{1}{Gm\mu} \frac{C_{\deform}}{(Gm)^{n_{\deform}}}
(Gm\omega)^{2n_{\deform}/3} \nn\\
&+ \frac{G}{\eta^{2}(Gm)^{2}}\frac{|\mu_{\txt{red}}|^{2}}{(Gm)^{2s}}
(Gm\omega)^{2s-2/3} \nn\\
&\left.
+ \frac{G C_{h}}{\eta^{2}} (Gm\omega)^{2n_{h}/3}
\right] \,.
\end{align}
One can obtain the gravitational waveform phase $\Psi(f)$ for the
dominant harmonic in the
Fourier domain from the relation~\cite{Tichy:1999pv}
\begin{equation}
\frac{d^2 \Psi}{d \omega^2} = 2 \frac{dE}{d\omega} \frac{dt}{dE}\,.
\end{equation}
By substituting Eqs.~\eqref{eq:mod-energy} and~\eqref{eq:mod-Edot}
into the above equation and integrating, we obtain
\begin{align}
\label{eq:mod-phase}
\Psi (f) &\sim  \frac{1}{\eta} (\pi G m f)^{-5/3} +
\frac{1}{\eta^{2} G\, m^{2}} \frac{|\mu_{1}\mu_{2}|}{(Gm)^{2s}}
(\pi G m f)^{(4s-5)/3} \nn \\
&\qquad \qquad+ \frac{1}{\eta^2 G\, m^{2}} \frac{C_\deform}{(Gm)^{n_\deform }}
(\pi G m f)^{(2n_\deform-5)/3} \nn\\
&\qquad\qquad + \frac{1}{\eta^3 G\, m^{2}} \frac{|\mu_\txt{red}|^2}{(Gm)^{2s}}
(\pi G m f)^{(6s-7)/3} \nn \\
&\qquad\qquad + \frac{G C_{h}}{\eta^3} (\pi G m f)^{(2n_h-5)/3}
\end{align}
for the dominant harmonic.
The leading term corresponds to the leading GR gravitational-wave
phase.  Each of the four correction terms arises from a unique
physical effect. We will number these effects as follows:
\begin{enumerate}
\item The correction proportional to $|\mu_{1}\mu_{2}|$ comes
  from the scalar pole-pole interaction modifying the binding energy
  and Kepler relation.
\item The correction proportional to $C_{\deform}$ comes from
  the metric deformation modifying the binding energy and Kepler
  relation.
\item The correction proportional to $|\mu_{\txt{red}}|^{2}$
  comes from the energy lost via scalar radiation.
\item The correction proportional to $C_{h}$ comes from the
  correction to the gravitational wave energy flux.
\end{enumerate}
In EDGB, correction 3 (scalar energy loss) dominates, giving a $-1$~pN
correction relative to GR~\cite{2012PhRvD..85f4022Y}, while in dCS all
the correction terms contribute at the same order, $+2$~pN relative to
GR~\cite{2013PhRvD..87h4058Y,Yagi:2012vf}.

\subsection{Mapping to post-Einsteinian parameters}
\label{sec:ppE}

The gravitational waveform phase in alternative theories of gravity
can be expressed using the so-called parametrized post-Einsteinian
(ppE) waveform phase $\Psi_\ppE(f)$ as~\cite{Yunes:2009ke}
\begin{equation}
\Psi_\ppE(f) = \Psi_\txt{GR}(f) + \beta_\ppE (\pi G\mathcal{M} f)^{b_\ppE}\,,
\end{equation}
where $\mathcal{M} = m \eta^{3/5}$ is the chirp mass. The correction
found in Eq.~\eqref{eq:mod-phase} can be mapped to the ppE waveform phase
above.

The four corrections enumerated above correspond to
\begin{subequations}
    \label{eq:b-ppE-all}
  \begin{align}
    \label{eq:b-ppE-1}
    b^{(1)}_\ppE &= (4s-5)/3  \\
    \label{eq:b-ppE-2}
    b^{(2)}_\ppE &= (2n_\deform-5)/3 \\
    \label{eq:b-ppE-3}
    b^{(3)}_\ppE &= (6s-7)/3 \\
    \label{eq:b-ppE-4}
    b^{(4)}_\ppE &= (2n_h-5)/3 \,.
  \end{align}
\end{subequations}
We may also extract the $\beta_\ppE$ parameters from
Eq.~\eqref{eq:mod-phase} after converting to $\mathcal{M}$. Each one
of the four $\beta$'s should be proportional to
a power of $\ell$. In this paper we have developed the scalings for
corrections 1 and 3 (proportional to $|\mu_{1}\mu_{2}|$ and
$|\mu_{\txt{red}}|^{2}$, respectively). For corrections 2 and 4 we can
only go so far as to say
\begin{subequations}
  \label{eq:beta-ppE-all}
  \begin{align}
    \label{eq:beta-ppE-2}
    \beta^{(2)}_{\ppE} &\sim \frac{1}{\eta^{1+2n_{\deform}/5}G \, m^{2}}
    \frac{C_{\deform}}{(Gm)^{n_{\deform}}} \\
    \label{eq:beta-ppE-4}
    \beta^{(4)}_{\ppE} &\sim \frac{G C_{h}}{\eta^{2+2n_{h}/5}} \,.
  \end{align}
For correction 1 we may go farther by using the scaling from
Sec.~\ref{sec:multipole-estimates-scos}. For simplicity, we are
focusing on the case with $s=t$, and $|\epsilon|=0$. This gives
\begin{equation}
  \label{eq:beta-ppE-1}
\lcheck
  \beta^{(1)}_{\ppE} \sim \frac{\ell^{2}}{\eta^{1+4s/5}(G m)^{2}}
  \left(
\frac{\ell^{2}}{R_{1}R_{2}}
  \right)^{\wp}
  \left(
    \frac{R_{1}R_{2}}{(Gm)^{2}}
  \right)^{s} (C_{1}C_{2})^{r} \,.
\end{equation}
Finally for correction 3 we may use the scaling estimates for
$\mu_{\txt{red}}$ from Sec.~\ref{sec:multipole-estimates-bin}. As we
saw, the scaling of $\mu_{\txt{red}}$ is controlled by the difference
$s-\wp$, and we gave four examples in Eq.~\eqref{eq:mu-red-s-p-all}
for $s-\wp=-1,0,+1,+2$. For these same four values we have
\begin{align}
  \label{eq:beta-ppE-3-s-p--1}
  \lcheck
  \beta^{(3)}_{\ppE} &\sim \frac{1}{\eta^{(12+6\wp)/5}}
  \left( \frac{\ell}{Gm} \right)^{2+2\wp}
  \left( \frac{\delta m}{m} \right)^{2}& (s-\wp=-1) \\
  \label{eq:beta-ppE-3-s-p-0}
  \lcheck
  \beta^{(3)}_{\ppE} &\sim \frac{1}{\eta^{(8+6\wp)/5}}
  \left( \frac{\ell}{Gm} \right)^{2+2\wp}
  \left( \frac{\delta m}{m} \right)^{2}& (s-\wp=0) \\
  \label{eq:beta-ppE-3-s-p-1}
  \lcheck
  \beta^{(3)}_{\ppE} &\sim \frac{1}{\eta^{(4+6\wp)/5}}
  \left( \frac{\ell}{Gm} \right)^{2+2\wp}
  & (s-\wp=+1)\\
  \label{eq:beta-ppE-3-s-p-2}
  \lcheck
  \beta^{(3)}_{\ppE} &\sim \frac{1}{\eta^{(10+6\wp)/5}}
  \left( \frac{\ell}{Gm} \right)^{2+2\wp}
  \left( \frac{\delta m}{m} \right)^{2}\,. & (s-\wp=+2)
\end{align}
\end{subequations}

We can compare these results with some already present in the
literature. Reference~\cite{2012PhRvD..85f4022Y} computed the
correction due to effect 3 (scalar energy flux correction) for black
hole binaries in both EDGB and dCS. For EDGB they found
$b_{\ppE}=-7/3$ and $\beta_{\ppE} \sim \zeta_{3} \eta^{-18/5} (\delta
m/m)^{2}$. Here, from Eq.~\eqref{eq:b-ppE-3} we find the same value of
$b_{\ppE}$. We use the $s-\wp=-1$ result
[Eq.~\eqref{eq:beta-ppE-3-s-p--1}] which also agrees with the result
of~\cite{2012PhRvD..85f4022Y} once we make the identification of
$\zeta_{3}\sim (\ell/Gm)^{4}$ (this is in agreement with their
definition of $\zeta_{3}$ and our earlier identification of
$\alpha_{3}\sim\mpl\ell^{2}$).

For dCS, Ref.~\cite{2012PhRvD..85f4022Y} found $b_{\ppE}=-1/3$ and
$\beta_{\ppE} \sim \zeta_{4} \eta^{-14/5} \bar{\bf{\Delta}}^{2}$ where
their $\bar{\bf{\Delta}}=\chi_1 \hat{\bf{S}}_{1} m_{2}/m \minusonetwo$
is a dimensionless vector encoding some combination of the spins of
the black holes (with $0\le\chi_{A}\le 1$ the dimensionless spin of
body A). This combination has the property that
$\bar{\bf{\Delta}}^{2} \to (\delta m/m)^{2}$ in the limit of coaligned
maximal spins. While our analysis ignored spin (we only considered
$|\epsilon|=0$) we have agreement on $b_{\ppE}$ and the scaling of
$\beta_{\ppE}$ with $\eta$. A more thorough analysis would capture the
spin dependence in $\bar{\bf{\Delta}}$, which at least gives the same
$(\delta m/m)$ dependence we find in the coaligned extremal spin
limit. Again we need to identify $\zeta_{4}\sim(\ell/Gm)^{4}$ (and
this again agrees with their definition of $\zeta_{4}$ and our earlier
identification of $\alpha_{4}\sim\mpl\ell^{2}$).

\section{Bounds estimate}
\label{sec:bounds-est}

In this section we estimate the bounds that could be placed on
$\ell$, from measurements of pericenter precession in pulsar binaries
(Sec.~\ref{sec:pericenter-precession}) and from gravitational wave
measurements (Sec.~\ref{sec:gw-sig}).

\subsection{Pericenter precession bounds}
\label{sec:peric-prec-bounds}

We again consider a pulsar binary system. In order to bound $\ell$ in
some given theory [with values given for $(|\epsilon|,d,r,\ell_{\NS},\ell_{\BH})$] requires
a high-quality timing solution with several post-Keplerian (pK) 
parameters well constrained~\cite{lrr-2008-8}. In this case the
pericenter precession is measured within some variance
$\sigma^{2}_{\avg{\dot\omega}}$ (and covariant with other timing
parameters, which we ignore here for simplicity). A proper constraint
on $\ell$ would require forming more timing solutions with $\ell$ a
free parameter, including the additional precession given in
Eq.~\eqref{eq:avg-omega-dot}. However here we can make a simple
estimate of the bounds which could be placed.

A simple estimate comes from ascribing all the variance $\sigma^{2}$
(we now drop the subscript) to the additional
precession in Eq.~\eqref{eq:avg-omega-dot}. We combine this with the
scaling estimate given in Eq.~\eqref{eq:avg-dot-omega-scaling}: change
the LHS to the ratio $|\sigma/\avg{\dot{\omega}}|$ and the
scaling to an inequality. This can be solved, for a given theory's
parameters and system's parameters, for a bound on $\ell$.
Specifically, we will have the scaling (now taking $s=t$)
\begin{equation}
\label{eq:ell-bound-est-pericenter}
\ell^{2+2\wp} \lesssim
\frac{|\sigma|}{\avg{\dot{\omega}}}
\frac{ GmG\mu R_{1}^{\wp}R_{2}^{\wp} }{(4s+1)!! C_{1}^{r}C_{2}^{r}}
\left[
\frac{(Gm)^{2}}{R_{1}R_{2}}
\right]^{s}
v^{2(1-2s)} \,.
\end{equation}

\begin{figure}[tbp]
  \centering
  \includegraphics[width=\columnwidth]{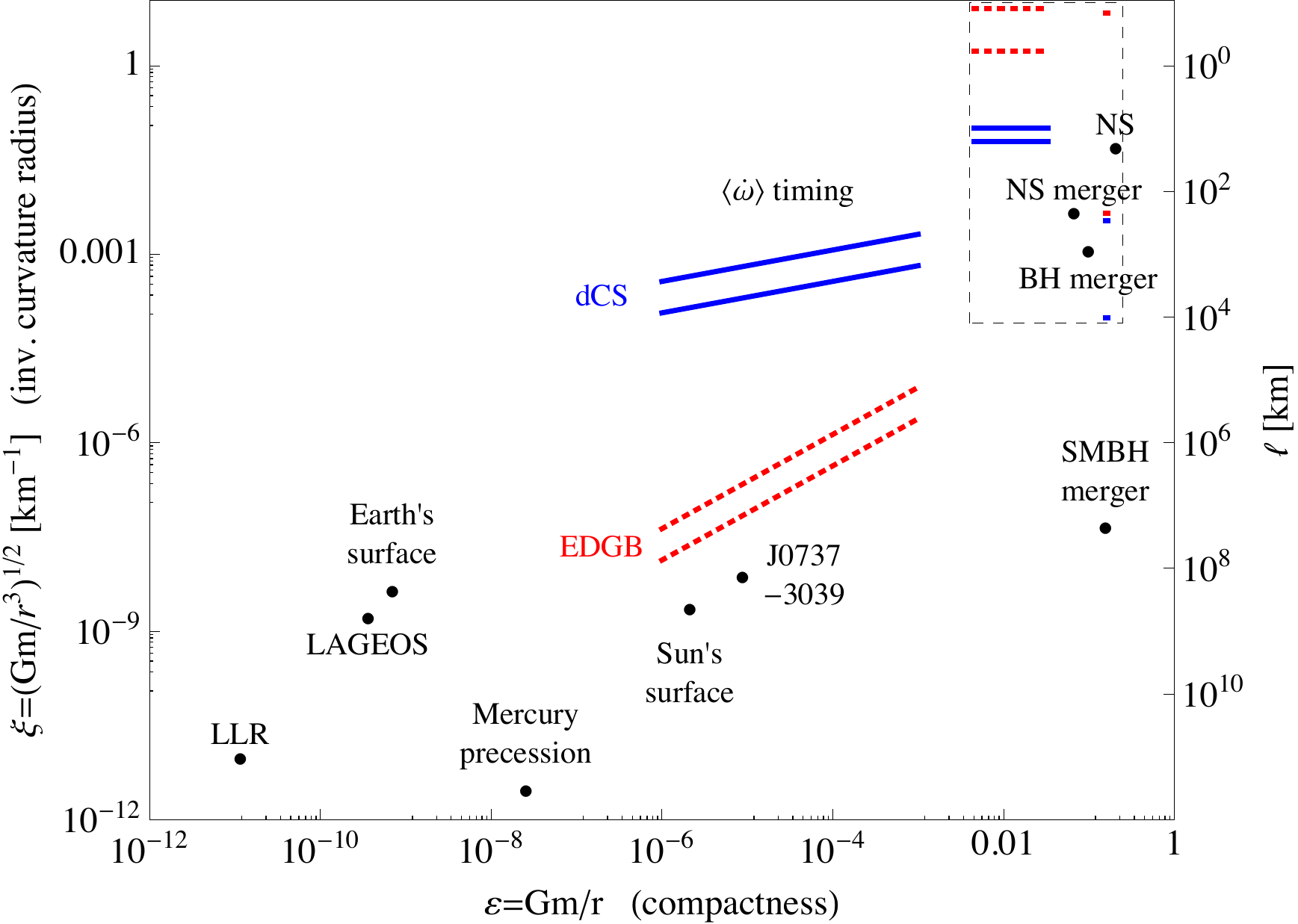}
  \caption{
    Estimated bounds on $\ell$.
    Upward sloping curves are estimates from pericenter precession
    coming from Eq.~\eqref{eq:ell-bound-est-pericenter}.
    The vertical axis at right is the length scale $\ell$ of the bound.
    The horizontal axis gives the compactness $(Gm/a)$ of a binary
    which yields a bound.
    Solid (blue online) curves correspond to bounds for dCS, while dotted
    (red online) curves correspond to EDGB.
    The lower curve for each theory is the estimate for
    $|\sigma/\avg{\dot{\omega}}|\sim 1$ while the
    upper curve is for
    $|\sigma/\avg{\dot{\omega}}|\sim 10^{-2}$.
    The dashed region is expanded in Fig.~\ref{fig:est-bounds-beta} to
    show estimated bounds from gravitational waves.
    \label{fig:est-bounds-all}}
\end{figure}

Some examples of such estimated bounds are plotted in
Fig.~\ref{fig:est-bounds-all}. On the horizontal axis we have the
dimensionless compactness of a pulsar binary system [related to the
orbital velocity in Eq.~\eqref{eq:avg-dot-omega-scaling} via the
leading-order Kepler relation $v^{2}=Gm/a=\varepsilon$]. On the (right,
inverted) vertical axis is the estimated bound on $\ell$. Values of
$\ell$ shorter than (therefore up in the plot) the plotted curves
would be allowed, while values greater (therefore down in the plot)
would be ruled out. The pericenter
precession estimates appear as sloped lines---clearly, larger values
of $Gm/a$ would produce better constraints on $\ell$.

The solid (blue online) curves correspond to the parameters of dCS while the
dotted (red online) curves correspond to the parameters of EDGB. The
lower curve of each pair corresponds to a value of
$|\sigma/\avg{\dot{\omega}}|\sim 1$ while the upper curve of each pair
corresponds to $|\sigma/\avg{\dot{\omega}}|\sim 10^{-2}$. Naturally a
better measurement of $\avg{\dot{\omega}}$ leads to a better
constraint on $\ell$. To generate each curve we used a fiducial
NS-NS system (so $s=t$) with masses $m_{1}=1.4 M_{\odot}=m_{2}$ and
radii $R_{1}=10 \text{km}=R_{2}$. The different slopes arise from the
different values of $\ell_{\NS}$ in each theory.
Equation~\eqref{eq:ell-bound-est-pericenter} does not include any spin
effects, which are required in dCS and likely required in EDGB for
NSs. To attempt to include these spin effects, we suppressed
$\avg{\dot{\omega}}$ by $\chi_{1}\chi_{2}$ for dCS, and took a
fiducial spin period of 300~ms~\cite{FaucherGiguere:2005ny} which gives
$\chi\approx 7\times 10^{-4}$. Similarly, since the scalar quadrupole
for a NS in EDGB is sourced at second order in spin, we suppressed the
pericenter precession in EDGB by $\chi_{1}^{2}\chi_{2}^{2}$ when
generating Fig.~\ref{fig:est-bounds-all}.

For comparison, Ref.~\cite{2011PhRvD..84l4033A} estimated a bound
$\ell_{\txt{dCS}} \lesssim 10^{8}$km from Solar System experiments. The
pulsar bounds here are estimated to be better by 4 or more orders of
magnitude.
The bound we have estimated here for dCS is consistent with the
calculation in Ref.~\cite{2013PhRvD..87h4058Y}.
However, this estimated bound must be interpreted with caution. Notice
in Fig.~\ref{fig:validity} that when $\ell\sim 30$km or larger, an
isolated neutron star will be in the large-modification
regime. However, the bounds estimated here were calculated with the
mass, radius, and multipole structure of the constituent neutron star.
While it should be safe to use the scaling of these properties in the
small-modification regime, it is not clear that these parameters scale
as assumed into the large-modification regime. This means these
estimated bounds may not be robust.

\subsection{Gravitational wave bounds}
\label{sec:gw-bounds}

\begin{figure}[tbp]
  \centering
  \includegraphics[width=\columnwidth]{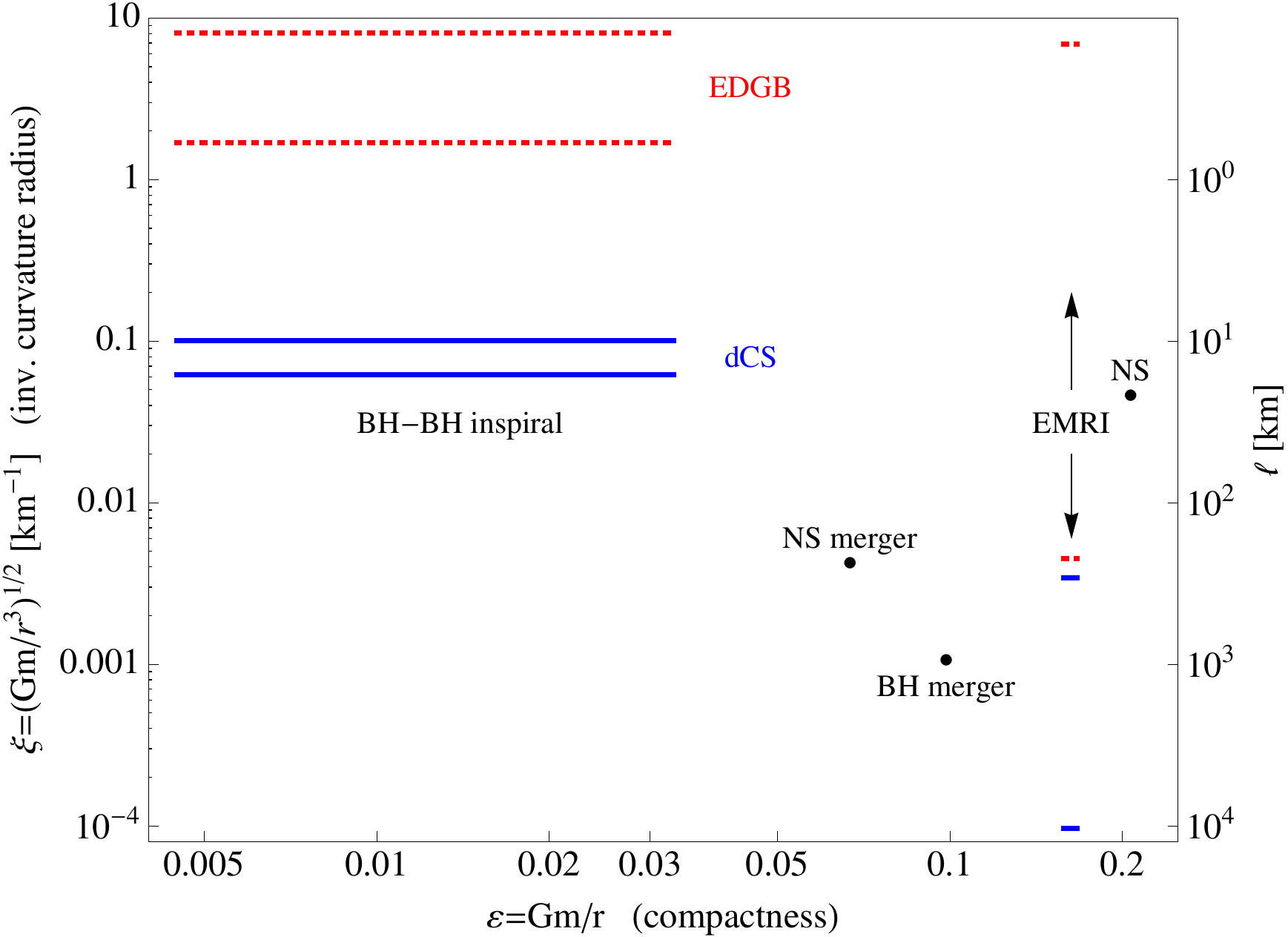}
  \caption{
    Estimated bounds on $\ell$ from gravitational wave
    measurements, assuming a detection at SNR 30 which is consistent
    with GR. This Figure is the dashed region
    within Fig.~\ref{fig:est-bounds-all}. Solid
    (blue online) lines correspond to dCS, dotted (red online) lines
    correspond to EDGB. Estimated bounds from a BH-BH inspiral in LIGO
    appear at left, with the horizontal extent of the line
    representing the range of frequencies in band. Estimates from an
    extreme mass-ratio inspiral (EMRI) detected in LISA with the small body being a black hole
    appear at right---they evolve through a narrower frequency
    range.
    \label{fig:est-bounds-beta}}
\end{figure}

We now return to a binary inspiral detected through gravitational
waves with some given signal/noise ratio (SNR). After a detection, to
properly bound $\ell$ in some given theory [with values given for
$(|\epsilon|,d,r,\ell_{\NS},\ell_{\BH})$] would involve integrating
against templates that include all the corrections in
Sec.~\ref{sec:gw-sig} and treating $\ell$ as a free
parameter. However, from the work of~\cite{Cornish:2011ys} we can
make a simple estimate of the bounds which could be placed. Their
Eq.~(20) estimates a bound
\newcommand{\udiff}[1]{\Delta u^{b^{#1}}}
\begin{equation}
\label{eq:Cornish-beta-estimate}
  |\beta|\lesssim \frac{3}{\SNR \udiff{}}
\end{equation}
where we have defined the shorthand $\Delta
u^{b}\equiv|u_{\min}^{b_{\ppE}}-u_{\max}^{b_{\ppE}}|$,
with $u=\pi G\mathcal{M}f=\eta^{3/5}v^{3}$, the min and max referring
to the frequency range where the signal is in band.
This can be directly converted into bounds on $\ell$ from either the
$\beta^{(1)}$ effect given in Eq.~\eqref{eq:beta-ppE-1} or the
$\beta^{(3)}$ effect given in
Eqs.~\eqref{eq:beta-ppE-3-s-p--1}-\eqref{eq:beta-ppE-3-s-p-2}.
For example, for the $\beta^{(1)}_{\ppE}$ bound we find [compare with
Eq.~\eqref{eq:ell-bound-est-pericenter}]
\begin{equation}
  \label{eq:ell-bound-beta-1}
  \ell^{2+2\wp} \lesssim \frac{3}{\SNR\udiff{(1)}}
  \frac{\eta^{+1+4s/5}}{C_{1}^{r}C_{2}^{r}}
  (Gm)^{2}R_{1}^{\wp}R_{2}^{\wp}
  \left[
\frac{(Gm)^{2}}{R_{1}R_{2}}
  \right]^{s} \,.
\end{equation}
Meanwhile, for $\beta^{(3)}_{\ppE}$ we find the simple expressions
\begin{subequations}
  \begin{align}
  \label{eq:ell-bound-beta-1-s-p--1}
    \ell^{2+2\wp} &\lesssim
    \frac{3 \eta^{(12+6\wp)/5}}{\SNR\udiff{(3)}}(Gm)^{2+2\wp}
    \left(
      \frac{m}{\delta m}
    \right)^{2} &(s-\wp=-1) \\
  \label{eq:ell-bound-beta-1-s-p-0}
    \ell^{2+2\wp} &\lesssim
    \frac{3 \eta^{(8+6\wp)/5}}{\SNR\udiff{(3)}}(Gm)^{2+2\wp}
    \left(
      \frac{m}{\delta m}
    \right)^{2} &(s-\wp=0) \\
  \label{eq:ell-bound-beta-1-s-p-1}
    \ell^{2+2\wp} &\lesssim
    \frac{3 \eta^{(4+6\wp)/5}}{\SNR\udiff{(3)}}(Gm)^{2+2\wp}
    &(s-\wp=+1) \\
  \label{eq:ell-bound-beta-1-s-p-2}
    \ell^{2+2\wp} &\lesssim
    \frac{3 \eta^{(10+6\wp)/5}}{\SNR\udiff{(3)}}(Gm)^{2+2\wp}
    \left(
      \frac{m}{\delta m}
    \right)^{2}\,. &(s-\wp=+2)
  \end{align}
\end{subequations}

To generate the estimated bounds in Fig.~\ref{fig:est-bounds-beta} we
considered two systems, two theories, and both $\beta^{(1)}$ and
$\beta^{(3)}$, for a product of eight estimated constraints. Each
constraint came from assuming an SNR 30 detection, i.e.~that $\beta$
could be bounded at the level of
\begin{equation*}
|\beta|\lesssim \frac{0.1}{\udiff{}} \,.
\end{equation*}
The two theories under consideration are dCS, represented as a solid
line (blue online), and EDGB, represented as a dotted line (red
online) with the parameters given in Table~\ref{table:params}. The two
systems under consideration were a stellar mass BH-BH inspiral detected in LIGO, and
an EMRI detected in LISA where the small object is a BH. For the
stellar mass BH-BH
system we took fiducial parameters $m_{1}=10 M_{\odot}$ and $m_{2}=11
M_{\odot}$. We represent the frequency
range during which the inspiral is in band as the horizontal extent of
the line, roughly 20--400Hz. The BH-BH LIGO estimates appear in the
left of the figure.
For the EMRI system we took parameters $m_{1}=10^{6} M_{\odot}$,
$m_{2}=10 M_{\odot}$. The frequency range ends when the small body
plunges, at a compactness of $\varepsilon=1/6$, and starts 1 year
before plunge. Since an EMRI evolves very slowly, the frequency range
is quite narrow. These estimates appear in the right of the figure.
We took all BHs to be rapidly spinning so that there is no spin
suppression in dCS.

The same expression [Eq.~\eqref{eq:ell-bound-beta-1}] is used for all
of the $\beta^{(1)}$ constraints, just with different parameters and
different values of $b_{\ppE}$. Further the $\beta^{(3)}$ expressions
depend on the difference $s-\wp$, which differs for the combinations
of theories and systems. For dCS we have $s-\wp=0$, given by
Eq.~\eqref{eq:ell-bound-beta-1-s-p-0}. In EDGB, we $s-\wp=-1$ for a
BH-BH binary, given by Eq.~\eqref{eq:ell-bound-beta-1-s-p--1} [for a
NS-NS (not considered here) we would have $s-\wp=+1$]. For dCS we have
$b_{\ppE}=-1/3$ in all cases, whereas in EDGB we must use
$b^{(1)}_{\ppE}=-5/3$ and $b^{(3)}_{\ppE}=-7/3$. For the combination
of dCS and stellar mass BHs, the constraint coming from $\beta^{(1)}$
is stronger than the $\beta^{(3)}$ constraint (and so is higher up in
Fig.~\ref{fig:est-bounds-beta}). For all other combinations of
theories and systems, the situation is reversed and the $\beta^{(3)}$
bound is stronger.

Now, let us compare the estimated bounds from GW observations shown in
Fig.~\ref{fig:est-bounds-beta} with those from previous
works~\cite{2012PhRvD..85f4022Y,2012PhRvD..86h1504Y,Yagi:2012vf,Canizares:2012is}. The
estimated GW bounds on EDGB found here are slightly stronger than the
previously estimated bounds
from~\cite{2012PhRvD..85f4022Y,2012PhRvD..86h1504Y}. This is because
in this paper, we did not take correlations among model parameters
into account, whereas constraints from previous works are either based
on a Bayesian~\cite{2012PhRvD..85f4022Y,Cornish:2011ys} or a Fisher
analysis~\cite{2012PhRvD..86h1504Y}, which weakens the bounds due to
parameter correlations. However, as an order of magnitude estimate,
our results are consistent with
Refs.~\cite{2012PhRvD..85f4022Y,2012PhRvD..86h1504Y}.

For the bounds
on dCS, the one from a BH-BH inspiral is of the same order as the one
estimated in~\cite{Yagi:2012vf}, where the authors performed a Fisher
analysis and included other corrections that we do not take into
account in this paper. On the other hand, the EMRI bound is larger
than the one estimated in~\cite{Canizares:2012is} by more than one
order of magnitude. This is because the latter considered a dCS
correction due to the modification in the gravitomagnetic component
of the metric to linear order in the BH spin, which is of higher pN
order than the one considered in this paper (2pN effect).  Finally let
us reiterate the caveat raised at the end of
Sec.~\ref{sec:peric-prec-bounds}, now relevant for the estimated dCS
bound coming from EMRIs. This analysis used the mass and multipole
structure of the small black hole, but the multipoles may not scale as
expected into the strong-modification regime. Therefore the dCS EMRI
bound, which is not smaller than kilometer scale, may not be robust.

\section{Conclusion}
\label{sec:conclusion}

In this paper, we have connected observables---pulsar binary
pericenter precession, and binary gravitational wave phase---to the
physical structure of the theory in a large class of models which
includes Einstein-dilaton-Gauss-Bonnet and dynamical Chern-Simons. We
have estimated upper limits which one would find from observations
consistent with general relativity, from both pericenter precession
and gravitational waves. Both bounds are expected to be orders of
magnitude better than Solar System bounds, with gravitational waves
being between 1 and 8 orders of magnitude better than pulsar timing
(depending on the compactness of the pulsar binary). The typical
length scale for gravitational-wave bounds is estimated to be
$\ell\lesssim 10$km.

To perform these calculations, we have parametrized the nonminimal
interaction Lagrangian $\calL_{\INT}$ in terms of the new length scale
$\ell$, which acts as a coupling parameter, and in terms of the
presence/absence of parity violation $|\epsilon|$, the number of
derivatives in the interaction $d$, and the number of curvature
invariants $r$.
We also parametrized the scalar field sourced by compact bodies in
terms of the leading nonvanishing multipole number, $\ell_{\NS}$ and
$\ell_{\BH}$. We have estimated the multipole moments $\mu^{Q}$ from
scaling arguments, which agree with asymptotic matching to
strong-field calculations for known examples.
We treat the compact objects with post-Newtonian theory by describing
them as effective point particles with scalar hair. This allows us to
derive an effective scalar multipole-multipole interaction Lagrangian
$L_{\INT}[{\bf x}_{1},{\bf x}_{2}]$ and compute the pericenter
precession in a compact binary system.
We also computed the scalar multipole moments of the binary and thus
the radiative scalar field and energy loss.

We used the stationary phase approximation to compute the modification
to the gravitational wave phase, by combining the modified binding
energy, modified Kepler relation, and the corrected energy loss. This
we connect to parameters in the parametrized post-Einsteinian (ppE)
framework: the $b$ and $\beta$ parameters arising from four distinct
effects. These effects are from (i) the conservative scalar
interaction, (ii) the conservative metric deformation, (iii) the energy
lost in scalar radiation, and (iv) the correction to the energy lost in
gravitational waves.

From both the pericenter precession and gravitational wave
calculations we can estimate the bound that would be placed from
observations consistent with general relativity
(Fig.~\ref{fig:est-bounds-all}). These bounds are estimated to be
orders of magnitude better than Solar System constraints. The
gravitational wave bounds are between 1 and 3 orders of magnitude
better than those arising from highly precessing pulsar binary systems
(depending on the compactness of the system). The typical length scale
for gravitational-wave bounds is estimated to be $\ell\lesssim 10$s of
km.

However, these estimates must be interpreted cautiously. For some
range of $\ell$, a constituent compact body may be in the
large-correction regime of the theory even though the binary is in the
small-correction regime, so these bounds may not be robust.


In this work we have captured only the scalar effects in the
theories we considered. Though we parametrized some of the metric
effects, we were unable to say how they scaled and thus unable to
determine what kinds of bounds they could provide. Another possible
extension of this present work would be to determine how these metric
effects generically scale.

Even within the realm of scalar-tensor theories, there are a variety
of phenomena available which do not fall into the framework we have
presented. Specifically, if the scalar multipole moments of compact
objects significantly change over an orbital or radiation-reaction
time scale, then we cannot ignore their time derivatives as we have
throughout this work. The most well-known example of this phenomenon
is the so-called spontaneous scalarization~\cite{Damour:1993hw} which
has been extensively numerically
investigated~\cite{Healy:2011ef,Barausse:2012da,Palenzuela:2013hsa,Shibata:2013pra,Berti:2013gfa}. This
effect is related to the presence of a carefully chosen potential for
the scalar field, whereas we have taken the potential to vanish in
this paper.

Further, this work has only considered theories with a metric and a massless
scalar field. Although this includes a large class of theories, there
are many more types of theories which are not included. Some examples
are bimetric theories~\cite{Hassan:2011zd} and tensor-vector or tensor-vector-scalar~\cite{Bekenstein:2004ne}
theories. In particular, here we cannot capture any Lorentz-violating
effects present in a theory such as Einstein-\AE{}ther~\cite{Jacobson:2008aj}. Additionally,
the multipole structure in tensor-vector and especially
Lorentz-violating theories is likely much richer than what is possible
in a scalar-tensor theory. Extending this work to include these
effects is a straightforward avenue for future investigation.

\acknowledgments
This work was inspired by a discussion at the \emph{Gravitational Wave
  Tests of Alternative Theories of Gravity in the Advanced Detector
  Era} workshop at MSU, and sharpened by discussions at a long-term
workshop at the Yukawa Institute for Theoretical Physics at Kyoto
University. We would like to acknowledge the organizers of both
workshops for creating programs that encouraged interaction and for
their hospitality during the programs. We would like to thank
E.~Barausse, E.~Berti, \'E.~Flanagan, L.~Lehner, and N.~Yunes for useful discussion and helpful
comments.
L.C.S. acknowledges that support for this work was
provided by the National Aeronautics and Space Administration through
Einstein Postdoctoral Fellowship Award No.~PF2-130101 issued by the
Chandra X-ray Observatory Center, which is operated by the Smithsonian
Astrophysical Observatory for and on behalf of the National
Aeronautics Space Administration under Contract No.~NAS8-03060,
and further acknowledges support from NSF Grant No.~PHY-1068541. K.Y.
acknowledges support from NSF Grant No.~PHY-1114374, NSF CAREER Grant
No.~PHY-1250636, and NASA Grant No.~NNX11AI49G.

\appendix

\section{Relation between $\ell$ and cutoff}
\label{sec:cutoff}
In this work we have parametrized the nonminimal interaction term
$\calL_{\INT}$ through the length scale $\ell$. However, in the EFT
framework, this term should arise from integrating out some unknown
physics above an energy scale $\Lambda$, the cutoff of this effective
theory. With the physics above this energy scale unknown, naturalness
arguments are typically invoked to estimate the sizes of
irrelevant/marginal/relevant operators in the action. Here we can
recast the action given in Eq.~\eqref{eq:action} in terms of
$\Lambda$. This is useful for estimating quantum corrections,
though in this paper we have treated everything classically.

First, we must rewrite the Einstein-Hilbert (E-H) term in terms of a
canonically normalized field variable. The Ricci scalar is expanded as
\begin{equation}
R\sim (\pd h)^{2} + \pd^{2}h\,.
\end{equation}
The E-H term, $\calL_{\txt{EH}}\sim
\mpl^{2}R$, becomes canonical by absorbing one power of $\mpl$ into
$h$, i.e.~defining
\begin{equation}
  h^{\txt{can}} \equiv \mpl h\,,
\end{equation}
so that $\calL_{\txt{EH}}\sim (\pd h^{\txt{can}})^{2}$. Here
$h^{\txt{can}}$ has canonical length dimensions,
i.e.~$[h^{\txt{can}}]=[L]^{-1}$.

Performing this redefinition in the interaction Lagrangian gives
\begin{equation}
\calL_{\INT} \sim \frac{(\mpl\ell)\ell^{\wp}}{\mpl^{2r}} \theta
T[\epsilon^{0,1}, \pd^{d}, (\pd h^{\txt{can}})^{2r}] \,.
\end{equation}
Now that all fields in this term are canonically normalized, from
naturalness we will argue that the coefficient of this term should be
an $\mathcal{O}(1)$ number times an inverse power of the cutoff
$\Lambda$. Specifically, for dimensional correctness we must have
\begin{equation}
\calL_{\INT} \sim \frac{c_{1}}{\Lambda^{4r+d-3}} \theta
T[\epsilon^{0,1},\pd^{d},(\pd h^{\txt{can}})^{2r}]
\end{equation}
where $c_{1}$ is some coefficient of order unity.
This alternate parametrization in terms of $c_{1}/\Lambda^{4r+d-3}$
is just as valid as the one in terms of $(\mpl\ell)\ell^{\wp}$.
Immediately we have the relation between $\Lambda$ and $\ell$,
\begin{equation}
\Lambda\sim c_{2}
\left[
\mpl^{(2r-1)} \ell^{-(2r+d-2)}
\right]^{1/(4r+d-3)}
\end{equation}
with $c_{2}$ another order-unity coefficient. We see that the cutoff
is parametrically between the inverse length $1/\ell$ and the Planck
scale $\mpl$.

For example, in both EDGB and dCS we have the scaling
\begin{equation}
\Lambda  \sim c_{2} \mpl^{3/5} \ell^{-2/5}\,.
\end{equation}
In terms of energies,
\begin{equation*}
  1 \text{km}^{-1} \approx 2\times 10^{-10} \text{eV}\,.
\end{equation*}
This gives an order of magnitude for the cutoff energy
\begin{equation*}
  \Lambda \sim c_{2} \cdot 3 \text{TeV}
  \left(
\frac{\ell}{10\text{km}}
  \right)^{-2/5}\,.
\end{equation*}
From Fig.~\ref{fig:validity} we see that for the end point of a
NS-NS merger to be within the regime of validity we should have
$\ell\lesssim 400$km in dCS and EDGB. This translates, with the above,
into having a cutoff at least as large as $\Lambda \gtrsim
0.7$TeV. If we want the structure of a NS to be within the
small-correction regime, then we see again from Fig.~\ref{fig:validity}
that we want $\ell\lesssim 30$km. This corresponds to a cutoff at
least as large as $\Lambda \gtrsim 2$TeV.

\clearpage{}


\bibliographystyle{apsrev4-1}
\bibliography{scalar-NS-effect}

\end{document}